\documentclass[10pt,journal,compsoc]{IEEEtran}
\ifCLASSOPTIONcompsoc
  \usepackage[nocompress]{cite}
\else
  \usepackage{cite}
\fi
\ifCLASSINFOpdf
  \usepackage[pdftex]{graphicx}
  \graphicspath{{fig/}}
  \DeclareGraphicsExtensions{.pdf,.jpeg,.png}
\else
  \usepackage[dvips]{graphicx}
  \graphicspath{{fig/}}
  \DeclareGraphicsExtensions{.eps}
\fi
\usepackage{amsmath}
\usepackage{algorithmic}

\usepackage{array}

\ifCLASSOPTIONcompsoc
  \usepackage[caption=false,font=footnotesize,labelfont=sf,textfont=sf]{subfig}
\else
  \usepackage[caption=false,font=footnotesize]{subfig}
\fi
\begin{document}
%
\title{Delivering Scientific Influence Analysis As a Service on Research Grants Repository}
%
%
%
%

\author{Yuming Wang,
        Yanbo Long,
        Lai Tu,
        Ling Liu, \IEEEmembership{Fellow, IEEE}
\IEEEcompsocitemizethanks{\IEEEcompsocthanksitem Yuming Wang, Yanbo Long and Lai Tu are with the School of Electronic Information and Communications, Huazhong University of Science and Technology, Wuhan 430074, Hubei, China.\protect\\
E-mail: ymwang@mail.hust.edu.cn
\IEEEcompsocthanksitem Ling Liu is with the School of Computer Science, College of Computing, Georgia Institute of Technology, Atlanta, GA 30332, USA.}
}

%
%

\markboth{}%
{Yuming Wang \MakeLowercase{\textit{et al.}}: GImpact}
%



\IEEEtitleabstractindextext{%
\begin{abstract}
Research grants have played an important role in seeding and promoting fundamental research projects worldwide. There is a growing demand for developing and delivering scientific influence analysis as a service on research grant repositories. Such analysis can provide insight on how research grants help foster new research collaborations, encourage cross-organizational collaborations, influence new research trends, and identify technical leadership. This paper presents the design and development of a grants-based scientific influence analysis service, coined as {\sc GImpact}. It takes a graph-theoretic approach to design and develop large scale scientific influence analysis over a large research-grant repository with three original contributions. First, we mine the grant database to identify and extract important features for grants influence analysis and represent such features using graph theoretic models. For example, we extract an institution graph and multiple associated aspect-based collaboration graphs, including a discipline graph and a keyword graph. Second, we introduce self-influence and co-influence algorithms to compute two types of collaboration relationship scores based on the number of grants and the types of grants for institutions. We compute the self-influence scores to reflect the grant based research collaborations among institutions and compute multiple co-influence scores to model the various types of cross-institution collaboration relationships in terms of disciplines and subject areas. Third, we compute the overall scientific influence score for every pair of institutions by introducing a weighted sum of the self-influence score and the multiple co-influence scores and conduct an influence-based clustering analysis. We evaluate {\sc GImpact} using a real grant database, consisting of 2512 institutions and their grants received over a period of 14 years. Our experimental results show that the {\sc GImpact} influence analysis approach can effectively identify the grant-based research collaboration groups and provide valuable insight on an in-depth understanding of the scientific influence of research grants on research programs, institution leadership, and future collaboration opportunities.
\end{abstract}

\begin{IEEEkeywords}
Research Grant database, Scientific influence, Graph mining, Graph clustering.
\end{IEEEkeywords}}

\maketitle

\IEEEdisplaynontitleabstractindextext

%
\IEEEpeerreviewmaketitle

\newcommand{\me}{\mathrm{e}}
\newcommand{\mi}{\mathrm{i}}
\newcommand{\dif}{\mathrm{d}}
\newcommand{\abs}[1]{\lvert#1\rvert}
\newcommand{\norm}[1]{\lVert#1\rVert}
\newcommand{\avg}{\mathop{\mathrm{avg}}}
\newcommand{\argmin}{\mathop{\mathrm{argmin}}}
\newcommand{\argmax}{\mathop{\mathrm{argmax}}}
\renewcommand{\vec}[1]{\mbox{\boldmath$#1$}}
\renewcommand{\algorithmicrequire}{\textbf{Input:}}
\renewcommand{\algorithmicensure}{\textbf{Output:}}
\renewcommand{\algorithmicrepeat}{\textbf{Repeat}}
\renewcommand{\algorithmicuntil}{\textbf{Until}}
\renewcommand{\algorithmicreturn}{\textbf{Return}}
\newtheorem{example}{Example}
\newtheorem{definition}{Definition}

\IEEEraisesectionheading{\section{Introduction}}

Research grants from governments and industry have played an important role in seeding and fostering fundamental and cutting-edge research projects, resulting in many research innovations and scientific discoveries. However, existing scientific influence analysis services to date have mainly centered on evaluating the impact factor of a journal or a conference based on citation counts. The first proposal for the Journal Impact Factor was introduced by E. Garfield in 1955\cite{garfield1955citation} to evaluate the influence of journals. It has been developed for more than 60 years\cite{garfield2006history} and is still widely used today.
In 2005, Hirsch\cite{hirsch2005indexcite, hirsch2007does} proposed the h-index to measure the influence of an individual researcher by combining both quantify (the number of publications) and quality (the count of citations). Several variant indices have been proposed to further enhance h-index, such as g-index\cite{egghe2006theory} and hg-index\cite{alonso2009hg} by adding one or more new attributes into indices or changing the way of processing citation counts. Such publication citation-based impact factor has been used by many research labs and academic institutions as one factor to evaluate the scholarly achievement of a researcher. Google scholar is a popular service for such purpose.

As big data and cloud computing become ubiquitous, there is a growing demand for developing large scale scientific influence analysis on research grants repositories and delivering such analysis as a service. In contrast to the publication citation counts, the scientific influence analysis on research grants can provide insight on how research grants help foster new research collaborations, encourage cross-organizational collaborations, influence new research trends, and identify technical leadership. For example, by examining the research grant repository over a certain period of time, it can reveal a number of interesting perspectives: in which subject areas academic institutions and industry researchers collaborate by means of cross-organization projects, the type of influence that research grants have on prioritizing certain research subjects over the others, on the research trends, and the leadership in different research subject areas. Analysis of research grants data may also reveal the specific research subject areas that are on-demand or on the priority-list by governments or industry. However, very few research efforts have been engaged on grants based scientific influence analysis using statistical methods~\cite{boyack2003indicator, jacob2011impact}.

In this paper, we develop a graph-theoretic approach to mine a research grants repository for large scale grants-based influence analysis, coined as {\sc GImpact}, and our design and development goal is to deliver {\sc GImpact} as a service with three original contributions.
First,
we mine a large scale grant database to identify and extract important features for grant-based scientific influence analysis and represent such features using graph theoretic models. For example, we can extract features to analyze research collaborations among different individual researchers and/or among different institutions by constructing a grant based researcher collaboration graph or an institution collaboration graph. For each of such graphs, we can again associate multiple aspect-based collaboration graphs that further characterize the grant-based collaborations through different aspects of collaboration, such as the disciplines identified by funding agencies of the research grants or the subject areas or keywords. Due to the space constraint, in this paper, we focus on the scientific influence analysis of grants on cross-institution collaboration. Thus, we construct an institution collaboration graph with institutions as vertices and joint grants between a pair of institutions as an edge. Similarly, we construct associated aspect-based collaboration graphs as additional features to enrich our influence analysis on cross-institution collaborations, such as a discipline graph and a keyword graph. The discipline graph has the disciplines as vertices and the grant based relationships between disciplines as edges with edge weighted by the total number of grants that are relevant to both disciplines. The keyword graph reflects the relationship among subject areas in the context of grants, and has the subject keywords as vertices and an edge between a pair of keywords if both keywords are covered by some grant(s), weighted by the total number of grants that cover both of the keywords.
Second,
we develop graph-theoretical algorithms to compute the collaboration relationship score between a pair of institutions based on their grant data to reflects two types of influences: self-influence and co-influence. We compute self-influence scores for each pair of institutions in terms of joint-grants based collaboration relationship, taking into account also the traversal reachability on the institution collaboration graph. We also compute the co-influence scores for each pair of institutions by incorporating each associated aspect-based collaboration graph. For example, if one institution is reachable from another institution through the graph traversal between the institution graph and one of its associated collaboration graphs, such as the discipline graph or the keywords graph, we will compute their co-influence score based on the statistical properties of all possible graph-traversal paths among the two institutions.
Third,
we compute the overall scientific influence scores by integrating the self-influence score and the multiple co-influence scores for each pair of institutions and conduct a scientific influence based clustering analysis on the institution graph by partitioning the institution collaboration graph into $K$ clusters, with $K$ as one of the service application interface parameters. The {\sc GImpact} approach presents a general purpose scientific influence analysis as a service framework and a suite of graph-theoretic influence computation algorithms that are capable of mining large scale grant data repositories with an easy-to-use API.
We evaluate {\sc GImpact} using a real grant database, consisting of 2512 institutions and their grants received over a period of 14 years. Our experimental results show that the {\sc GImpact} influence analysis approach can effectively identify the grant-based research collaboration groups and provide valuable insight and an in-depth understanding of the scientific influence of research grants on research programs, institution leadership, and future collaboration opportunities in different research subject areas.
\section{Overview}

\subsection{Research Grants Dataset}

The dataset for the study is obtained from the Social Sciences Management Databases of Chinese Universities (SMDB), which consists of all projects in Humanities and Social Sciences from the Ministry of Education, China from the period of the year 2005 to the year 2018. Table~\ref{tab:grants} shows research grant samples, Table~\ref{tab:institutions}, ~\ref{tab:disciplines}, ~\ref{tab:keywords} show institution samples, discipline samples and keyword samples respectively. Table~\ref{tab:stat} shows basic statistical characteristics of the dataset.

	{\small
		\begin{table}[!htbp]
			\renewcommand{\arraystretch}{1.3}
			\caption{The Research Grant Samples}
			\label{tab:grants}
			\centering
			\begin{tabular}{
				p{0.05\textwidth}<{\raggedright}
				p{0.115\textwidth}<{\raggedright}
				p{0.115\textwidth}<{\raggedright}
				p{0.11\textwidth}<{\raggedright}}
				\hline
				Record & Institution  & Discipline             & Keyword            \\
				\hline
				R01    & CUFE, SHUFE  & D63040, D79071         & K06, K11           \\
				R02    & SHUFE, SWUFE & D63044, D79071, D84074 & K08, K12, K13      \\
				R03    & BNU, FUDAN   & D79071, D81030, D88031 & K03, K08           \\
				R04    & RUC          & D7907340               & K01, K09           \\
				R05    & FUDAN        & D7907340               & K07, K05, K02, K04 \\
				R06    & CUFE         & D7907340               & K01, K10           \\
				\hline
			\end{tabular}
		\end{table}
	}

	{\small
		\begin{table}[!htbp]
			\renewcommand{\arraystretch}{1.3}
			\caption{The Institution Samples}
			\label{tab:institutions}
			\centering
			\begin{tabular}[b]{
				p{0.08\textwidth}<{\raggedright}
				p{0.36\textwidth}<{\raggedright}}
				\hline
				Institution & InstitutionName                                    \\
				\hline
				PKU         & Peking University                                  \\
				RUC         & Renmin University of China                         \\
				BNU         & Beijing Normal University                          \\
				CUFE        & Central University of Finance and Economics        \\
				CUPL        & China University of Political Science and Law      \\
				FUDAN       & Fudan University                                   \\
				ECNU        & East China Normal University                       \\
				SHUFE       & Shanghai University of Finance and Economics       \\
				ECUPL       & East China University of Political Science and Law \\
				WHU         & Wuhan University                                   \\
				SWUFE       & Southwestern University of Finance and Economics   \\
				SWUPL       & Southwest University of Political Science and Law  \\
				\hline
			\end{tabular}
		\end{table}
	}

	{\small
		\begin{table}[!htbp]
			\renewcommand{\arraystretch}{1.3}
			\caption{The Discipline Samples}
			\label{tab:disciplines}
			\centering
			\begin{tabular}[b]{
				p{0.08\textwidth}<{\raggedright}
				p{0.36\textwidth}<{\raggedright}}
				\hline
				Discipline & DisciplineName                         \\
				\hline
				D190       & Psychology                             \\
				D630       & Management                             \\
				D63040     & Enterprise Management                  \\
				D63044     & Public Management                      \\
				D740       & Linguistics                            \\
				D790       & Economics                              \\
				D79071     & Finance                                \\
				D7907340   & Financial Markets                      \\
				D81030     & Administration                         \\
				D820       & Law                                    \\
				D84074     & Labor Science                          \\
				D870       & Library, Information and Documentation \\
				D88031     & Educational Economics                  \\
				\hline
			\end{tabular}
		\end{table}
	}

	{\small
		\begin{table}[!htbp]
			\renewcommand{\arraystretch}{1.3}
			\caption{The Keyword Samples}
			\label{tab:keywords}
			\centering
			\begin{tabular}[b]{
				p{0.08\textwidth}<{\raggedright}
				p{0.36\textwidth}<{\raggedright}}
				\hline
				Keyword & KeywordName                 \\
				\hline
				K01     & Economic Cycle              \\
				K02     & Monetary Assets             \\
				K03     & Compulsory Education        \\
				K04     & International Pricing Power \\
				K05     & Bulk Goods                  \\
				K06     & Tax Policy                  \\
				K07     & Derivatives                 \\
				K08     & Financial                   \\
				K09     & Asset Pricing               \\
				K10     & Capital Market              \\
				K11     & Venture Capital             \\
				k12     & Supply Mechanism            \\
				k13     & Labor Force                 \\
				\hline
			\end{tabular}
		\end{table}
	}

	{\small
		\begin{table}[!htbp]
			\renewcommand{\arraystretch}{1.3}
			\caption{The Basic Statistical Characteristics}
			\label{tab:stat}
			\centering
			\begin{tabular}{
				p{0.34\textwidth}<{\raggedright}
				p{0.08\textwidth}<{\raggedright}}
				\hline
				Characteristics                              & Number      \\
				\hline
				\# of Research Grants                        & 334068      \\
				\# of Institutions                           & 2512        \\
				\# of Disciplines                            & 1569        \\
				\# of Keywords                               & 20097       \\
				\# of Institutions per Grant (Min, Avg, Max) & 1, 1.74, 10 \\
				\# of Disciplines per Grant (Min, Avg, Max)  & 1, 1.19, 5  \\
				\# of Keywords per Grant (Min, Avg, Max)     & 1, 7.87, 19 \\
				\hline
			\end{tabular}
		\end{table}
	}

Raw data is plain-text records in the database managed by a relational DBMS. In order to perform the proposed scientific influence analysis on the grant database, we need to perform feature extractions and convert the relational tables of grant records in plain text format into graph representations. For example, each record contains a collection of attributes, such as institutions involved, disciplines related and keywords associated, and so forth.
We show an example fragment of the grant record samples in Table~\ref{tab:grants}. For record R01, we can learn that institution CUFE and SHUFE have direct grant collaboration. The related discipline areas are D63040 and D79071, and the associated keywords are K06 and K11.
If we focus on analyzing the scientific influence of grants on research collaboration among institutions through subject areas captured by disciplines and keywords, then we can extract features from the grant database by modeling each grant by a selection of attributes, such as the institutions, the disciplines and the keywords, then we can formulate this projected version of the grant database as $\mathcal{G}$ with a collection of triples, each of the format $\mathcal{(I, D, K) \in \mathcal{G}}$, where $\mathcal{I}$ is the institution collection, $\mathcal{D}$ is the discipline collection, $\mathcal{K}$ is the keyword collection.
If the main focus of our scientific influence analysis is on institution collaboration, then we construct the institution graph first with institutions as vertices and joint grants between a pair of institutions as an edge weighted by the number of joint grants. For each additional attributes, we will construct an aspect-based collaboration graph, such as a discipline graph and a keyword graph, to highlight the relationship among different values of the attributes, and enrich our influence analysis on cross-institution collaborations. For each of the specific attributes, we can construct a graph by extracting the relationship features between the same type of attributes. Although each aspect-based collaboration graph is homogeneous in nature, the entire collection of graphs are heterogeneous with one primary attribute as the main collaboration graph for influence analysis and other attributes as the additional aspect of collaborations to capture the different aspects of collaboration relationships that are important to characterize the influence between the vertices in the main collaboration graph, i.e., institutions, in our case. The discipline graph has the disciplines as vertices and the grant based relationships between disciplines as edges with edge weighted by the total number of grants that are relevant to both disciplines. The keyword graph reflects the relationship among subject areas in the context of grants, and has the subject keywords as vertices and an edge between a pair of keywords if both keywords are covered by some grant(s), weighted by the total number of grants that cover both of the keywords.

We would also like to note that the techniques developed in our {\sc GImpact} is generic and one can choose other primary attributes instead of the institution, such as researcher who are the PI or co-PIs of a grant, making institutions as one aspect-based collaboration graph. Due to the space constraint, in this paper we focus on showcase our approach by conducting the scientific influence analysis of grants on cross-institution collaboration. We use two example attributes, disciplines and keywords, to illustrate the selection of attributes to extract features to represent different collaboration aspects in terms of graphs.

\begin{definition}[Research Grants Network]
	A research grants network is a heterogeneous information network and defined by an undirected graph $G = (V, E)$ where $V$ is the set of vertices of heterogeneous types, representing attributes of research grants, such as institution, discipline, and keyword, and $E$ is the set of edges denoting the heterogeneous relationships between a pair of vertices of homogeneous types, such as institution-institution, discipline-discipline, keyword-keyword.
\end{definition}

\begin{figure}[!htbp]
	\centering
	\includegraphics[width = 0.3\textwidth]{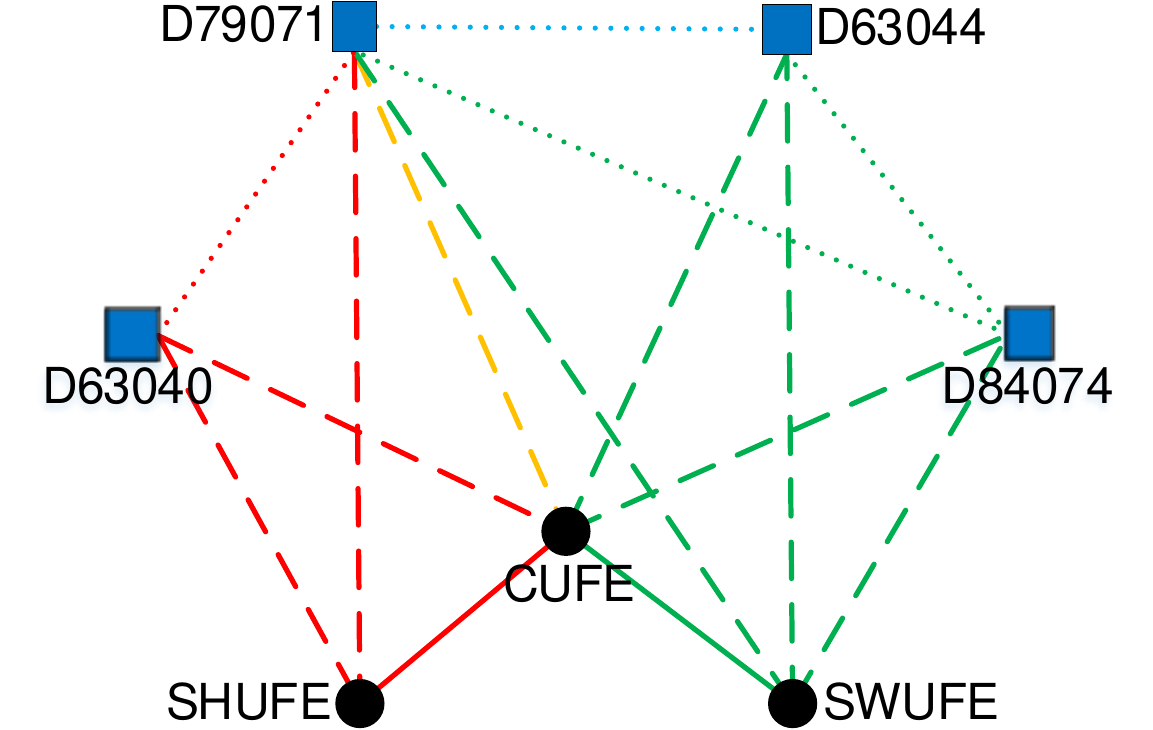}
	\caption{An illustration of the Research Grants Network}
	\label{fig:rgn-simple-example}
\end{figure}

Consider the research grant samples in Table~\ref{tab:grants}. From record R01, we can extract homogeneous links of CUFE-SHUFE, D63040-D79071 directly from the grant database and thus expressed in solid lines, and the heterogeneous links of CUFE-D63040, CUFE-D79071, SHUFE-D63040, and SHUFE-D79071, and thus expressed in dotted lines. Similarly, from record R02, we extract homogeneous links of SHUFE-SWUFE, D63044-D79071, D63044-D84074 and D79071-D84074 and heterogeneous links of SHUFE-D63044, SHUFE-D79071, SHUFE-D84074, SWUFE-D63044, SWUFE-D79071, and SWUFE-D84074.
Figure~\ref{fig:rgn-simple-example} shows an illustrating example of the research grants network built from record R01 and record R02.
It consists of two types of attributes (vertices): institutions (black circle) and disciplines (blue square) and three types of relationships (edges): institution-institution (solid line), discipline-discipline (dot line), and institution-discipline (dash line). The edges from record R01 are marked as red, the edges from record R02 are marked as green, and the edges from both record R01 and record R02 are marked as yellow.

By building a heterogeneous research grants network using multiple homogeneous networks directly from the grant database of plain-text grant records, we can perform {\sc GImpact} based scientific influence analysis on the multiple homogeneous graphs together and find both direct and indirect collaboration relationships among vertices of the primary collaboration graph, such as the institution graph and the relationships of CUFE-SHUFE, SHUFE-SWUFE, and CUFE-SHUFE-SWUFE in terms of not only joint grants but also the grant based scientific influence through mining all the homogeneous graphs collectively as a whole. As a byproduct, we can also learn about relationships between an institution and its associated aspects of collaborations, e.g., SHUFE took part in the disciplines of D63040, D63044, D79071, and D84074. We can learn hidden relationships that are not observable from the plain-text grant records through simple summarization techniques.

\subsection{Related Work}
The design and development of {\sc GImpact} are inspired primarily by social network analysis research efforts in the last decade.
Social network analysis promotes community detection~\cite{newman2004finding, fortunato2010community, khan2017network} and social influence computation~\cite{wang2012influence, arif2012scientific, bergsma2014learning, jiang2017measuring, giudice2018new, ma2008mining, zhou2013social}. Most of existing social network analysis techniques focus on the single network of homogeneous vertices with homogeneous links, such as a social network of people with friendships among people without explicitly modeling different types of links, which constrains the social influence analysis to be at the superficial social network connection specific friendships. Also, most of the existing social network influence analysis is based on the co-authorship using DBLP dataset. To our best knowledge, none of the prior work has explored scientific influence analysis on a large scale grants databases.

Existing research efforts on research grants data repositories are limited. \cite{boyack2003indicator} is the first to study the impact of governmental funding on the publication counts and their citation counts from research programs at the National Institute on Aging (NIA), aiming at improving the quality of funded research. \cite{jacob2011impact}  evaluates the impact of receiving NIH grants on the publication by comparing the impact of receiving an NIH grant on subsequent publications and citations with publications and citations from those with unsuccessful grant applications on standard research grants of R01s, showing the insignificant difference between these two groups in terms of both publication counts and citation counts. We argue that scientific influence analysis on the innovation of research programs, institution leadership, and future collaboration opportunities can be more useful indicators for grant impact evaluation than only based on publication count and citation count.

\subsection{Problem Statement}
The first problem we intent to address in the development of {\sc GImpact} is to develop graph-theoretic and statistical methods to compute indirect grant collaboration relationships among institutions based on those observable features captured in the grant database in terms of grant records.

Consider the research grant samples in Table~\ref{tab:grants}. From record R01, we can find that institution CUFE and SHUFE jointly applied for a grant in the disciplines of D63040 and D79071. From record R02, institution SHUFE and SWUFE jointly applied for a grant in the discipline of D63044, D79071, and D84074. Although CUFE and SWUFE did not apply for a grant jointly, CUFE and SWUFE both applied for a grant with SHUFE. Furthermore, CUFE and SWUFE both applied for a grant in the discipline of D79071.
Thus, only based on the joint grant information to conduct grant based scientific influence analysis may lead to some biased or inaccurate results. We argue that a comprehensive scientific influence analysis on a grant repository should take into account of not only direct relationship that can be obtained from the grant database records but also the many types of indirect relationships among institutions that have contributed to the research initiatives and research projects in the same or related disciplines and on the same or similar topic keywords. We argue that measuring grant based scientific influence across institutions should consider both direct and indirect collaborations in the context of grant data. Thus, it is important to extract features that are representing different collaboration aspects in addition to the grant data on institutions. For example, disciplines and keywords are important attributes that reflect the collaboration aspects of different institutions. Furthermore, the relationships between an institution and its associated disciplines in the grant disciplines graph and its associated topic keywords in the grant keywords graph are highly relevant as well.

The second problem we propose to tackle in {\sc GImpact} is to develop statistical mining algorithms to compute two types of influence measures: self-influence and co-influence. The self-influence refers to the influence score that is computed based only on the graph traversal information in a primary collaboration graph of homogeneous vertices, such as the institution graph in which vertices have edges between them if they have joint grants. The graph traversal on the joint grants based institution graph will capture the indirect relationship among institutions that have indirect grant-based collaboration relationships.
The co-influence refers to the influence score that is computed based on both the graph traversal information on a primary collaboration graph and its multiple associated aspect-based collaboration graphs. The graph traversal on this collection of homogeneous graphs will also capture the indirect relationship among institutions that have indirect grant-based collaboration relationships in terms of common disciplines or common topic keywords.

In the development of {\sc GImpact}, we attempt to answer two fundamental questions: (1) How to measure the overall scientific influence between any pair of institutions quantitatively; and (2) How to utilize the overall scientific influence scores to identify grant based institution clusters. To address the first question, we will compute the overall scientific influence score between any pair of institutions using a weighted sum of the self-influence score and the multiple co-influence scores. The overall scientific influence score reflects not only the collaboration patterns in the institution graph through direct and indirect joint grant relationships but also the collaboration patterns through common disciplines and common keywords as well as indirectly related disciplines in the discipline aspect graph and indirectly related topic keywords in the keyword aspect graph. To address the second question, we will develop scientific influence distance based graph clustering algorithm to partition the set of $n$ institutions, denoted by $I$, into $K$ disjoint clusters $I_j$ ($1 \leq j \leq K$), where $I = \bigcup^K_{j=1} I_j$ and $I_j \bigcap I_k = \emptyset$ for $\forall 1 \leq j, k \leq K, K \leq n$.  The clustering result should achieve a good balance between intra-cluster similarity, i.e., the vertices within one cluster should have close collaboration relationship and similar collaboration patterns, and inter-cluster similarity, i.e., the vertices in different clusters should have relatively loser collaboration relationship and dissimilar collaboration patterns.

The final outcome of {\sc GImpact} is the grant-based overall scientific influence for each given institution, which is represented by a ranked list of other institutions sorted by the influence score by this institution based on both the direct and indirect joint grants and the grants that are related directly or indirectly by common or similar disciplines and/or keywords.

\subsection{Solution Approach and Overall Framework}

Given a grant database of plain text records, the users of {\sc GImpact} is asked to identify the primary collaboration attribute, say institution, and the secondary attributes that can be modeled as different grant-aspect graphs, say disciplines and keywords. We then construct the corresponding research grants network in three steps. (1) We first extract all distinct institutions and their associated aspect attributes from the grant database and represent each institution with the selection of attributes, such as grant ID, institution ID, discipline IDs and keyword IDs. (2) We construct the institution graph, which contains homogeneous vertices of type institutions, and homogeneous edges between institutions if they have joint grants. We use {\sc GImpact} to learn self-influence collaboration patterns from direct and indirect joint-grant based collaborations. (3) We construct multiple grant-aspect graphs, each of which contains a set of homogeneous vertices of one attribute type and a set of homogeneous edges between two aspect vertices if they are reflected in a common grant, such as the two disciplines are in at least one grant, or the two keywords are appeared in at least one grant. Each of such grant-based aspect graphs will be used by {\sc GImpact} to perform the scientific influence analysis to learn co-influence collaboration patterns by exploring graph traversal across both the primary institution graph and the aspect graph by utilizing direct and indirect traversal paths.

\begin{definition}[Grant-based Institution Graph]
	A grant-based institution graph is a subgraph of $G=(V,E)$, and represented as $IG = (IV, IE)$, where $IG\subset V$ is the set of institutions, and $IE\subset E$ is the set of edges denoting the joint grant relationship between a pair of institutions. Let $N_{0}$ denote the total number of institutions in $I$, we have $N_{0} = |IV|$.
\end{definition}

\begin{definition}[Grant-based Aspect Graph]
	A grant-based aspect graph, denoted as $AG_i = (AV_i, AE_i)$, is a subgraph of $G=(V,E)$, corresponding to a grant-specific aspect attribute, such as disciplines, or keywords, where $AV_i$ is the set of distinct aspect attribute values, $AE_i$ is the set of edges denoting the direct relationship between two aspect values if they are covered by the same grant, reflecting the grant-specific aspect relevance, such as discipline relevancy in the discipline graph or the keyword relevancy in the keyword graph. $N_{i}$ denotes the total number of grant aspect specific vertices in $AE_i$, and $N_{i} = |AV_i|$.
\end{definition}

\begin{definition}[Grant-based Influence Graph]
	A grant-based influence graph is defined based on the institution graph $IG$ and a grant-based aspect graph $AG_i$, and is denoted as $FG_i = (IV, AV_i, AE_i, FE_i)$, where $IV\subset V$ is the set of institutions, $AV_i\subset V$ is the set of grant-based aspect vertices in the $i$-th aspect graph $AG_i$ ($1\leq i\leq N$), and $AE_i\subset E$ is the set of edges denoting the direct relationship between two distinct aspect attribute values, such as discipline relevancy and keyword relevancy, $FE_i$ is the set of edges, each connecting an institution vertex and an aspect vertex, denoting the direct relationship between an institution and its aspect attribute value, weighted by the \#grants this institution has with the same aspect attribute value, such as the same discipline in the discipline graph.
\end{definition}

\begin{figure*}[!htbp]
	\centering
	\includegraphics[width = \textwidth]{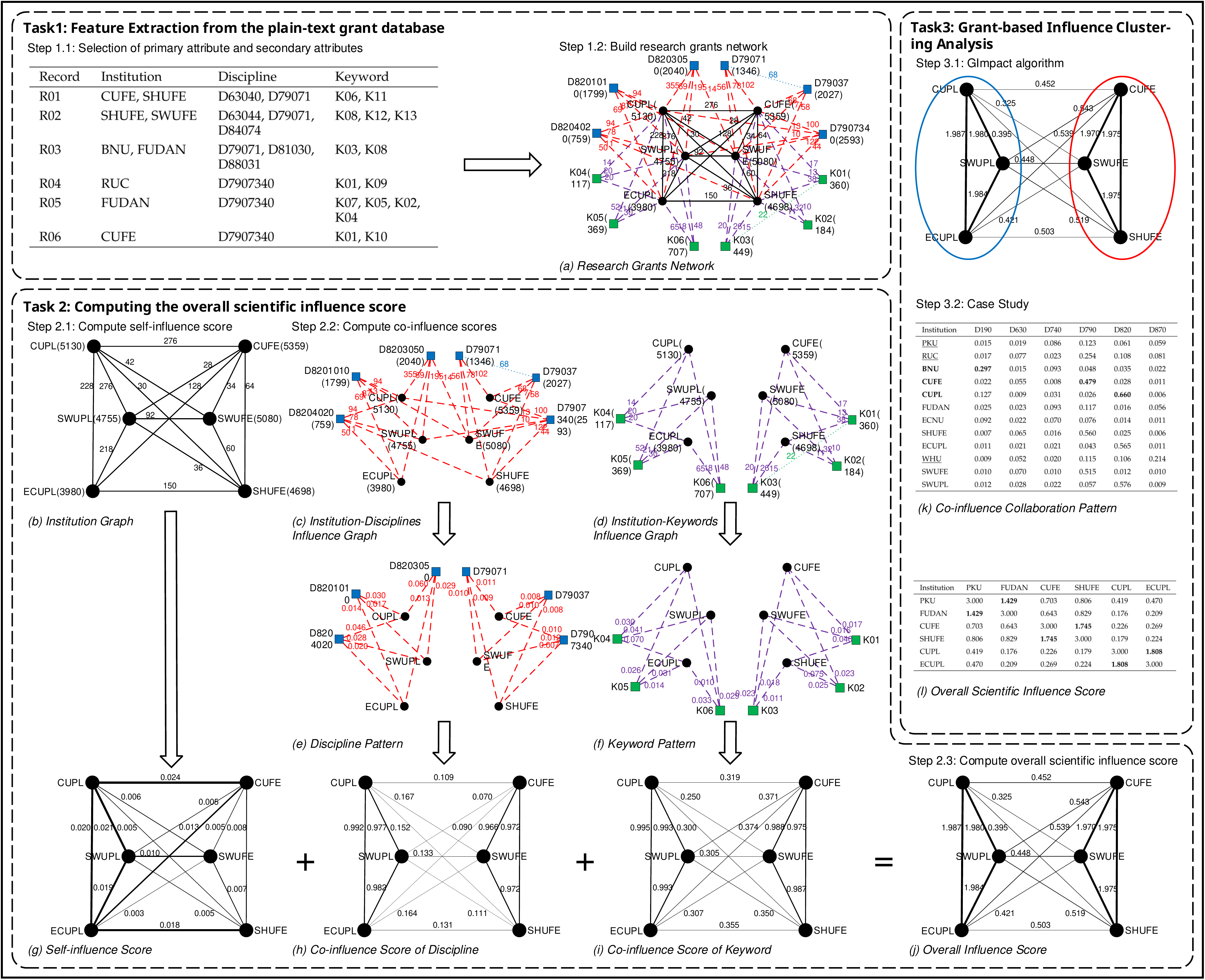}
	\caption{An Illustration of the workflow of the GImpact Influence Analysis Framework}
	\label{fig:framework}
\end{figure*}

Figure~\ref{fig:framework} provides an example workflow to illustrate the three main tasks of the {\sc GImpact} framework.
Figure~\ref{fig:framework}a shows an example fragment of the research grants network extracted from our grant database SMDB. It consists of three types of vertices: institutions (black circle), disciplines (blue square), and keywords (green square), as well as
five types of direct relationships (edges) extracted directly from the grant database plain-text records: institution-institution (black solid line), discipline-discipline (blue dotted line), keyword-keyword (green dotted line), institution-discipline (red dashed line), and institution-keyword (green dashed line). We decouple the research grants network in Figure~\ref{fig:framework}a into three sub-graphs: an institution graph in Figure~\ref{fig:framework}b, and two influence graphs: institution-disciplines influence graph in Figure~\ref{fig:framework}c and institution-keyword influence graph in Figure~\ref{fig:framework}d.
Black numbers in the bracket indicate \#grants of an institution, and black numbers on an institution-institution edge represent \#joint-grants.
Red numbers on the red dashed edge between an institution and a discipline denotes \#grants that the institution applied in that discipline. Also, purple number on the purple dashed edge between an institution and a keyword denotes \#grants that the institution applied to cover that keyword.
Blue numbers on a blue dotted edge represents \#grants covered by both disciplines. The green number on a green dotted edge represent \#grants that cover both keywords.
For ease of presentation, we removed the edges with weight less than 10.

{\bf Task 1: Feature Extraction from the plain-text grant database. } A user of our {\sc GImpact} service should first select the primary grant related attribute that he is interested in conducting grant based scientific influence analysis, such as institution, the lead PI or the entire PI and co-PIs team. Assuming we select the attribute institution in the grant record as our primary attribute for collaboration influence study. Then the user will need to select a set of secondary attributes as the collaboration aspects for the influence analysis. For example, common or similar disciplines and keywords can be indicators of research collaboration relevance. The selection of primary and secondary grant based attributes is always user-specific and influence task-specific.
%
Consider research grant samples in Table~\ref{tab:grants}. From record R01, we can learn the research domain is D63040 (Enterprise Management) and D79071 (Finance) from its disciplines, and the research content is about K06 (Tax Policy) and K11 (Venture Capital) from its topic keywords. By using these information selections, one can learn that the grant R01 is about tax policy and venture capital in the domain of enterprise management and finance. Such information can be leveraged to conduct research grants based influence analysis of institution CUFE or SHUFE respectively over the grant repository.
The outcome of the first task is the $N+1$ grant-based graphs, with an institution graph as the primary collaboration graph and $N$ grant-based aspect-specific influence graphs.

	{\bf Task 2: Computing the overall scientific influence score.\/} {\sc GImpact} performs this task by computing the self-influence score on the primary institution collaboration graph and the multiple co-influence scores on the $N$ grant aspect-specific influence graphs. We divide this step into three sub-tasks:
(1) Computing the self-influence score for every pair of vertices on the institution graph $IG$ based on the direct and indirect joint-grant relationships. We use a homogeneous influence spread model based on heat diffusion~\cite{ma2008mining} to compute the self-influence score. The result is a $N_{0} \times N_{0}$ matrix, denoted by $S_0$, with each entry representing the self-influence score of pair-wise institutions.
(2) Computing co-influence scores on each of the $N$ grant aspect-specific influence graphs {$FG_i$} (for $1\leq i\leq N$). We utilize the heterogeneous influence spread model~\cite{zhou2013social} to compute the co-influence scores. The result is a $N_{i} \times N_{i}$ matrix, denoted by $S_i$, each representing the co-influence that is spread between institutions via one grant-aspect specific attribute based influence graph such as disciplines graph. If we have $N$ number of aspect-specific influence graphs $FG$, thus we have $N$ co-influence score matrices produced in this task, denoted as $S_1, \cdots, S_N$.
(3) Finally, to obtain the overall influence score for each pair of institutions by integrating the self-influence score and the $N$ co-influence scores by deriving a weighted scheme and record the result in a $N_{0} \times N_{0}$ matrix, such as a weighted sum of the self-influence score with weight $\alpha$ and the $N$ co-influence scores with weights $\alpha, \omega_i, i = 1, \cdots, N$.

	{\bf Task 3: Grant-based Influence Clustering Analysis.\/} In addition to perform influence analysis using self-influence matrix $S_{0}$, co-influence matrices: $S_1, \cdots, S_N$, and the overall influence $S$, we want to utilize the overall influence scores as an influence based distance function to perform graph clustering analysis. For example, by using the K-Medoids clustering method \cite{kaufman1987clustering}, we design our influence based clustering algorithm {\sc GImpact} to partition all institutions into $K$ grant-based collaboration clusters.
Unlike conventional K-Medoids clustering method, we refine the centroid-based initialization function.
When assigning points into a cluster, we consider both the influence score to the centroid and the influence score to all points in the cluster.
Also, we select the new centroid by maximizing the intra-cluster influence similarity and minimizing the inter-cluster influence similarity.
\section{Scientific influence analysis}

In this section, we will discuss how to compute the overall scientific influence score $S$ for each pair of institutions.
We will first introduce the heat diffusion~\cite{ma2008mining} based general influence spread model to perform the influence spread process through graph traversal in a undirected graph.
Then we utilize the homogeneous influence spread model to compute self-influence score on the primary institution collaboration graph $IG$ and the heterogeneous influence spread model to compute multiple co-influence scores on the $N$ grant aspect-specific influence graphs $FG_i$.
Finally, we will integrate the self-influence score $S_0$ and the $N$ co-influence scores $S_1, \cdots, S_N$ into the overall scientific influence score $S$ with weights $\alpha, \omega_i, i = 1, \cdots, N$.

\subsection{General Influence Spread Model}

Inspired by the heat diffusion\cite{ma2008mining}, we developed our general influence spread model to perform the influence spread process through graph traversal.
Heat diffusion is a physical phenomenon that heat transfer from a hot object to a cold object. The heat diffusion phenomenon is very similar to the influence spread. For example, the extraordinary institution can be considered as the hot object, transfer heat to the cold object or spread influence to the moderate institution considered as the cold object.
Once an institution applied for a grant with other institutions. It means that this institution spread influence on other institutions. Also, if an institution applied for a grant in discipline, this institution spread influence on this discipline. The more institutions that applied for grants on this discipline, the more possibilities other institutions apply for grants on this discipline. Thus this institution influenced other institutions through this discipline.

Consider a undirected graph $G = (V, E)$, where $V$ is the vertex set and $E$ is the edge set. We use $N$ to represent the size of $V$, i.e. $N_v = |V|$. The vertex $v \in V$ can be considered as a object in the thermal system and the edge $(v_m, v_n) \in E$ can be considered as a heat transfer tunnel (influence path) between $v_m$ and $v_n$.
Suppose at time $t$, the amount of heat $\Delta H(v_m, v_n, t)$ that vertex $v_m$ received from $v_n$ during a period of $\Delta t$ should be proportional to the temperature $T_n(t)$ of the vertex $v_n$ at time $t$ and the probability $p_{mn}$ of the vertex $v_m$ receive heat from $v_n$. Based on the above assumptions, we can define $\Delta H(v_m, v_n, t) = p_{mn}T_n(t) \Delta t$.
As a result, the temperature change $\Delta T_m$ at vertex $v_m \in V$ between time $t + \Delta t$ and time $t$ is defined by the heat $\Delta H_R(v_m, v_n, t)$ it receives subtract the heat $\Delta H_S(v_n, v_m, t)$ it sends. This is formulated as

\begin{equation}
	\begin{split}
		\Delta T_m
		&= \alpha \sum_{v_n \in V, v_m \neq v_n} (\Delta H_R(v_m, v_n, t) - \Delta H_S(v_n, v_m, t)) \\
		&= \alpha \sum_{v_n \in V, v_m \neq v_n} (p_{mn}T_n(t) - p_{nm}T_m(t)) \Delta t \\
	\end{split}
	\label{eq:deltaT-scalar}
\end{equation}

where $\alpha$ is the heat conductivity or the influence spread coefficient. For ease of representation, we can express the above formulation into a matrix form:

\begin{equation}
	\frac{\Delta \vec{T}(t)}{\Delta t} = \alpha \vec{H} \vec{T}(t)
	\label{eq:deltaT-matrix}
\end{equation}

where

\begin{equation}
	H_{mn}= \begin{cases}
		\  p_{mn},                                 & m \neq n \\
		\  -\sum_{v_l \in V, v_l \neq v_m} p_{lm}, & m = n
	\end{cases}
	\label{eq:Hij-scalar}
\end{equation}

\begin{equation}
	\vec{T}(t) = \begin{bmatrix} T_1(t) & T_2(t) & \cdots & T_N(t)  \end{bmatrix} ^T
	\label{eq:Tt-scalar}
\end{equation}

where $\vec{H}$ is a $N_v \times N_v$ matrix, called the one-hop heat diffusion kernel or the one-hop influence spread kernel, as the heat diffusion only considers one-hop diffusion in the whole process. The value $H_{mn}$ when $m \neq n$ indicates the heat that vertex $v_m$ receives from its neighbor $v_n$, while the value $H_{mn}$ when $m = n$ indicates the heat that vertex $v_m$ sends to its all neighbors. The heat received should equal to the heat sent, therefore the sum of each row of the matrix should be 0.

In the limit $\Delta t \rightarrow 0$, this becomes

\begin{equation}
	\frac{\dif \vec{T}(t)}{\dif t} = \alpha \vec{H} \vec{T}(t)
	\label{eq:Tt}
\end{equation}

By solving this differential equation, we can get

\begin{equation}
	\vec{T}(t) = \me^{\alpha t \vec{H}} \vec{T}(0)
	\label{eq:Tt-solution}
\end{equation}

where $\vec{T}(0)$ is the initial heat distribution or the initial influence distribution. The matrix $\me^{\alpha t \vec{H}}$ is a $N \times N$ matrix, called the multi-hop heat diffusion kernel or multi-hop influence spread kernel, as the heat diffusion considers infinity times from the initial heat distribution. It can be extended as a Taylor series, where $\vec{I}$ is the identity matrix:

\begin{equation}
	\me^{\alpha t \vec{H}} = \vec{I} + \alpha t \vec{H} + \frac{\alpha^2t^2}{2!} \vec{H}^2  + \frac{\alpha^3t^3}{3!} \vec{H}^3 + \cdots.
	\label{eq:kernel-taylor-series}
\end{equation}

The multi-hop influence spread kernel $\me^{\alpha t \vec{H}}$ capture both direct and indirect influence paths between any two vertices in the undirected graph $G = (V, E)$. The influence spread coefficient $\alpha$ is a user-specific parameter, representing the speed of the influence spread process.

\subsection{Homogeneous Influence Spread Model}

The institution graph $IG = (IV, IE)$ is a homogeneous graph and only contains homogeneous edges between institutions. Based on the general influence spread model mentioned above, we only need to consider the homogeneous edges in the graph, i.e., $\vec{H}$ is a $N_0 \times N_0$ matrix. The probability $p_{mn}$ of the institution $iv_m \in IV$ receive influence from the institution $iv_n \in IV$ can be defined as

\begin{equation}
	p_{mn} = \begin{cases}
		\  \frac{w_{mn}}{\sqrt{w_m w_n}}, & (iv_m, iv_n) \in IE \\
		\  0,                             & \textit{otherwise}
	\end{cases}
	\label{eq:pij-ig}
\end{equation}

where $w_m$ (or $w_n$) is the weight of vertex $iv_m$ (or $iv_n$), e.g., \#grants of the institution $iv_m$ (or $iv_n$), and $w_{mn}$ is the weight of edge $(iv_m, iv_n)$, e.g., \#joint-grants between the institution $iv_m$ and the institution $iv_n$.

By defining the probability $p_{mn}$, we can compute the one-hop influence spread kernel $\vec{H}$ according to Eq.(\ref{eq:Hij-scalar}). Meanwhile, by giving a specific $\alpha$ and $t$, we can obtain the multi-hop influence spread kernel $\me^{\alpha t \vec{H}}$ according to Eq.(\ref{eq:kernel-taylor-series}) as well. The multi-hop influence spread kernel in homogeneous influence spread model is a $N_0 \times N_0$ matrix that capturing both direct and indirect collaboration relationships, i.e., self-influence collaboration patterns. Thus the self-influence score can be defined as $S_0 = \me^{\alpha t \vec{H}}$.

Figure~\ref{fig:framework}g shows an example of the self-influence score based on the institution graph of Figure~\ref{fig:framework}b, where both $\alpha$ and $t$ are set to 1.
The black number on black edge represents the self-influence score between institutions. Although the weight of edge CUPL-CUFE and the weight of edge CUPL-SWUPL are both 276 in Figure~\ref{fig:framework}b, the self-influence scores of edge CUPL-CUFE and edge CUPL-SWUPL are 0.024 and 0.021 in Figure~\ref{fig:framework}g respectively. It shows that the self-influence score can capture not only direct collaboration relationships but also indirect collaboration relationships.

\subsection{Heterogeneous Influence Spread Model}

The influence graph $FG_i = (IV, AV_i, AE_i, FE_i)$ is a heterogeneous graph and contains not only homogeneous edges between aspect attributes but also heterogeneous edges between institution and aspect attribute. Based on the general influence spread model mentioned above, we need to redefine the one-hop influence spread kernel $\vec{H}_i$ for heterogeneous graph~\cite{zhou2013social} to capture both homogeneous and heterogeneous edges. Thus we divide the one-hop influence spread kernel $\vec{H}_i$ into four parts:

\begin{equation}
	\vec{H}_i = \begin{bmatrix}
		\mathcal{(AA)} & \mathcal{(AI)} \\
		\mathcal{(IA)} & \mathcal{(II)}
	\end{bmatrix}
	\label{eq:Hij-scalar-dg}
\end{equation}

where $\mathcal{(AA)}$ is an $N_{i} \times N_{i}$ matrix with the similar definition as Eq.(\ref{eq:Hij-scalar}), representing the grant based relationship between aspect attributes, defined by Eq.(\ref{eq:aa-jk});
$\mathcal{(AI)}$ is an $N_{i} \times N_{0}$ matrix representing the scientific influence that
aspect attributes received from institutions, defined by Eq.(\ref{eq:ai-jk});
$\mathcal{(IA)}$ is an $N_{0} \times N_{i}$ matrix representing the scientific influence that institutions received from aspect attributes, defined by Eq.(\ref{eq:ia-jk});
and $\mathcal{(II)}$ is a $N_{0} \times N_{0}$ diagonal matrix to keep the sum of each row of the matrix to 0.

\begin{equation}
	\mathcal{(AI)}_{mn} = \begin{cases}
		\  \frac{w_{mn}}{\sum_{l=1}^{N_{i}}w_{ln}}, & (av_m, iv_n) \in FE_i \\
		\  0,                                       & \textit{otherwise}
	\end{cases}
	\label{eq:ia-jk}
\end{equation}

where $w_{mn}$ is the weight on influence path $(av_m, iv_n)$. $\mathcal{(AI)}_{mn}$ is defined by $w_{mn}$ normalized by the sum of weights on $(av_l, iv_n)$ for any $av_l$ in $AV_i$.

\begin{equation}
	\mathcal{(IA)}_{mn} = \begin{cases}
		\  \frac{w_{mn}}{\sum_{l=1}^{N_{0}}w_{ln}}, & (iv_m, av_n) \in FE_i \\
		\  0,                                       & \textit{otherwise}
	\end{cases}
	\label{eq:ai-jk}
\end{equation}

where $w_{mn}$ is the weight on influence path $(iv_m, av_n)$. $\mathcal{(IA)}_{mn}$ is defined by $w_{mn}$ normalized by the sum of weights on $(iv_l, av_n)$ for any $iv_l$ in $IV$.

\begin{equation}
	\mathcal{(AA)}_{mn} = \begin{cases}
		\ s_{mn},                                                                         & m \neq n \\
		\ -(\sum_{l=1}^{N_i} \mathcal{(AA)}_{ml} + \sum_{l=1}^{N_0} \mathcal{(AI)}_{ml}), & m = n
	\end{cases}
	\label{eq:aa-jk}
\end{equation}

\begin{equation}
	s_{mn} = \begin{cases}
		\  \frac{w_{mn}}{\sqrt{w_m w_n}}, & (av_m, av_n) \in AE_i \\
		\  0,                             & \textit{otherwise}
	\end{cases}
	\label{eq:pjk-dg}
\end{equation}

where $s_{mn}$ is the similarity between the aspect attribute $av_m$ and the aspect attribute $av_n$. $\sum_{l=1}^{N_i} \mathcal{(AA)}_{ml} + \sum_{l=1}^{N_0} \mathcal{(AI)}_{ml}$ summarizes the influence that the aspect attribute $av_m$ send to other aspect attributes and institutions.

For diagonal matrix $\mathcal{(II)}$, the diagonal entry is defined as $\mathcal{(II)}_{mm} = -\sum_{l=1}^{N_{i}} \mathcal{(IA)}_{ml}$ which summarizes the influence that the institution $iv_m \in IV$ send to other aspect attributes.

By defining the one-hop influence spread kernel $\vec{H}_i$ for heterogeneous network and giving a specific $\alpha$ and $t$, we can utilize Eq.(\ref{eq:kernel-taylor-series}) to obtain the multi-hop influence spread kernel $\me^{\alpha t \vec{H}_i}$. The multi-hop influence spread kernel in heterogeneous influence spread model is a $(N_{i}+N_{0}) \times (N_{i}+N_{0})$ matrix that capturing relationships between institutions and associated aspect attributes.
The $\mathcal{(AI)}$ part of the multi-hop influence spread kernel capture the scientific influence that aspect attributes received from institutions, i.e., co-influence collaboration patterns.

Figure~\ref{fig:framework}e and Figure~\ref{fig:framework}f show examples of co-influence collaboration patterns based on discipline influence graph of Figure~\ref{fig:framework}c and keyword influence graph of Figure~\ref{fig:framework}d respectively, where both $\alpha$ and $t$ are set to 1.
The red numbers on red dashed lines measure the scientific influence that disciplines received from institutions. The purple numbers on purple dashed lines measure the scientific influence that keywords received from institutions.
For ease of presentation, we removed the edges with weight less than 0.001.

\subsection{Co-Influence Score}

The $\mathcal{(AI)}$ part of the multi-hop influence spread kernel $\me^{\alpha t \vec{H}_i}$ represents the co-influence collaboration patterns in the associated influence graph. The co-influence collaboration pattern $\vec{p}_m$ of an institution $i_m \in IV$ is a vector and can be formulated as

\begin{equation}
	\vec{p}_m = \begin{bmatrix}
		\me^{\alpha t \vec{H}_i}(1, m + N_{i}) \\
		\me^{\alpha t \vec{H}_i}(2, m + N_{i}) \\
		\cdots                                 \\
		\me^{\alpha t \vec{H}_i}(N_{i}, m + N_{i})
	\end{bmatrix}
	\label{eq:behavior-pattern-scalar}
\end{equation}

The co-influence score is the similarity score of co-influence collaboration patterns between pair-wise institutions. The similarity of co-influence collaboration patterns should consider not only the angle difference but also the length difference between two vectors.
One prolific institution should have a longer length of co-influence collaboration vector because of more \#grants, while one ordinary institution should have a shorter length of co-influence collaboration vector because of fewer \#grants. Considering that one prolific institution and one ordinary institution may have little research collaboration relationship even if the angle or the proportion of participation is similar.
From this point on, we can define the co-influence score $S_{i}(m, n)$ between two institutions $iv_m, iv_n$ is

\begin{equation}
	S_{i}(m, n) = \left(
	\frac{\vec{p}_m \cdot \vec{p}_n}{\abs{\vec{p}_m} \abs{\vec{p}_n}}
	\right)\cdot\left(
	1 - \frac{\abs{\vec{p}_m - \vec{p}_n}}{\abs{\vec{p}_m} + \abs{\vec{p}_n}}
	\right)
	\label{eq:score-eg}
\end{equation}

Figure~\ref{fig:framework}h and Figure~\ref{fig:framework}i show examples of co-influence score based on discipline collaboration patterns of Figure~\ref{fig:framework}e and keyword collaboration patterns of Figure~\ref{fig:framework}f respectively.
The black numbers on black edges represent the co-influence score between aspect attributes.

From Figure~\ref{fig:framework}h and Figure~\ref{fig:framework}i, we can observe that the co-influence scores of discipline and keyword are similar but different.
It means that the aspect attributes of discipline and keyword are complementary to each other for research collaboration relevance.
Meanwhile, from Figure~\ref{fig:framework}g, the self-influence score and the co-influence score are quite different.
It means that the co-influence score contains different information from the self-influence score. Thus, combining them into an overall scientific influence score is of great significance.

\subsection{Overall Scientific Influence Measure}

The overall scientific influence score considers not only direct and indirect collaboration relationships between institutions, i.e., self-influence collaboration patterns, but also relationships between institutions and aspect attributes, i.e., co-influence collaboration patterns by integrating self-influence score and multiple co-influence scores.

By setting a given value for $t$, self-influence score $S_0(\alpha, t)$ can be rewritten as $S_0(\alpha)$, where $\alpha$ is the influence spread coefficient that can be used as the weight of self-influence score. The overall scientific influence score $S$ is defined as

\begin{equation}
	S = S_0(\alpha) + \sum_{i = 1}^N \omega_i S_{i}
	\label{eq:S-matrix}
\end{equation}

where $N$ is the number of influence graphs, $S_i$ is the $i^{th}$ co-influence score, $\omega_i$ is the weight for the $i^{th}$ co-influence score, and $\alpha  + \sum_{i=1}^N \omega_i = N + 1, \alpha \geq 0, \omega_i \geq 0, i = 1, \cdots, N.$ The scalar form can be written as

\begin{equation}
	S(m, n) = \begin{cases}
		\sum\limits_{l=0}^\infty \frac{\alpha^l t^l}{l!} H^l(m, n) +
		\sum\limits_{i=1}^N \omega_i S_{i}(m, n), & m \neq n \\
		N+1,                                      & m = n
	\end{cases}
	\label{eq:S-scalar}
\end{equation}

Figure~\ref{fig:framework}j shows an example of overall scientific influence scores based on self-influence score of Figure~\ref{fig:framework}g, discipline co-influence score of Figure~\ref{fig:framework}h and keyword co-influence score of Figure~\ref{fig:framework}i, where $\alpha$, $\omega_1$ and $\omega_2$ are all set to 1.
The black numbers on black edges represent the overall scientific influence between institutions.
\section{Graph Clustering Analysis}

In this section, we will introduce our grant-based influence clustering algorithm {\sc GImpact} to partition all institutions into $K$ grant-based collaboration clusters by utilizing the overall scientific influence score as an influence based distance function.
	{\sc GImpact} follows the conventional K-Medoids clustering algorithm \cite{kaufman1987clustering} and incorporate some new techniques on centroid initialization, vertex assignment and centroid update.
Distance metrics is an essential step for conventional K-Medoids clustering algorithm. A simple idea is to use the adjacency matrix. For a weighted undirected graph, the distance between two vertices is the sum of weights of each edge on the shortest path connecting them.
We argue that the overall scientific influence score can be more useful distance metrics than simple adjacency matrix.

\subsection{Purpose and Objective}

The purpose of the clustering analysis is to partition the institution set $I$ into $K$ disjoint clusters $I_j$, where $I = \bigcup^K_{j=1} I_j$ and $I_j \bigcap I_k = \emptyset$ for $\forall 1 \leq j, k \leq K, j \neq k$ to find close collaboration clusters. The objective of clustering analysis is to maximize intra-cluster influence score and minimize the inter-cluster influence score to achieve a good balance between the following two characteristics: (1) vertices within one cluster should have close collaboration relationship and similar collaboration patterns; (2) vertices in different clusters should have relatively loser collaboration relationship and dissimilar collaboration patterns.

\begin{definition}[Intra-Cluster Influence Score]
	The intra-cluster influence score is the average influence score of vertices in the same cluster to the centroid.
	For a group of disjoint clusters $\{I_1, I_2, \cdots, I_K\}$, $cv_j$ is the centroid of cluster $I_j$, the intra-cluster influence score of $I_j$ is defined as below:
	\begin{equation}
		S(cv_j, I_j) = \frac{1}{\abs{I_j}} \sum_{iv_l\in I_j} s(cv_j, iv_l)
		\label{eq:intra-score}
	\end{equation}
\end{definition}

\begin{definition}[Inter-Cluster Influences Score]
	The inter-cluster influence score is the average influence score of vertices in one cluster $I_j$ to the centroid $cv_k$ of another cluster $I_k$.
	For a group of disjoint clusters $\{I_1, I_2, \cdots, I_K\}$, $cv_j$ (or $cv_k$) is the centroid of cluster $I_j$ (or $I_k$), the inter-cluster influence score between $I_j$ and $I_k$ is defined as below:
	\begin{equation}
		\begin{split}
			S(I_j, I_k)
			&= \frac{1}{2} \left( S(cv_j, I_k) + S(cv_k, I_j) \right) \\
		\end{split}
		\label{eq:inter-score}
	\end{equation}
\end{definition}

Without loss of generality, a good centroid can represent the cluster. Moreover, only considering the sum of the centroid to vertices in the cluster will effectively reduce the time complexity from $O(N^2)$ to $O(N)$. It is essential for a large scale graph.

\subsection{Centroid Initialization}

Centroid Initialization is the first step for K-Medoids clustering algorithm. The purpose of centroid initialization is to obtain a collection of initial centroids $C^* = \{cv_1^0, cv_2^0, \cdots, cv_K^0\}$. Good initial centroids usually have good clustering results, while bad initial centroids are the opposite.
There are lots of different centroid initialization schema, such as DENCLUE \cite{hinneburg1998efficient} and K-Means++ \cite{arthur2007k}. The original K-Medoids clustering algorithm uses the random initialization schema. The idea of DENCLUE is to choose the local maximum of the density function as centroids. The idea of K-Means++ is to choose the first centroid at random and to choose remaining centroids with probability proportional to its squared distance from the closest existing center. We test different centroid initialization schemes and discuss their respective features.

\subsubsection{Random Initialization}

The $K$ centroids are completely randomly selected in this schema. The random result is the baseline of all other schemes.

\subsubsection{Top-K Degree Initialization}

Degree is the number of edges that a node has \cite{hanneman2005introduction}.
A vertex which has greater degrees has a local maximum of the number of neighbors. It can diffuse heat to as many vertices as possible. It is obviously better than a completely random selection in the data space.
We sort all vertices in the descending order of their degrees and select top-K vertices as the initial $K$ centroids.

\subsubsection{Top-K Density Initialization}

The density function of a vertex $iv_j \in I$ is the sum of all influence scores related to itself.

\begin{equation}
	D(iv_j) = \sum_{iv_l \in I, iv_j \neq iv_l} s(iv_j, iv_l)
	\label{eq:density-function}
\end{equation}

The larger the density value of a vertex, the faster the vertex can diffuse and receive influence. Compared to top-K degree schemes, this schema considers the weights of its edge.
Thus we can select top-K vertices by sorting vertices in the descending order of their density value as the initial $K$ centroids.

\subsubsection{Mixed Initialization}

Inspired by K-Medoids++ \cite{arthur2007k}, we develop a mixed centroid initialization schema. A good centroid should be far from all other centroids, but it should not be an outlier either.
Our initialization schema chooses the first centroid by the max density value of vertices, calculated by Eq.(\ref{eq:density-function}). Then we choose remaining centroids by considering both average influence score and maximum influence score from other existed centroids.
A new centroid with the smallest average influence score will make it as far as possible from other existing centroids. However, if the distance to other existed centroids is unevenly distributed, for example, this distance to one of the centroids is very close, and the distance to other centroids is very far, the average collaboration score may also be quite small.
A new centroid with the smallest maximum influence score will make it as far as possible from the nearest existing centroid. However, it may select an outlier in practice.
Based on above considerations, we define the mix score $M(iv_j)$ of a vertex $iv_j \in I$ as:

\begin{equation}
	\begin{split}
		M(iv_j) &= \frac{\sum_{iv_j \neq cv_k} S(iv_j, cv_k)}
		{\Delta S(iv_j) \cdot \abs{C^*}} \\
		&+ \Delta S(iv_j) \cdot \max_{iv_j \neq cv_k}{S(iv_j, cv_k)}, \\
		\Delta S(iv_j) &= \max\limits_{iv_j \neq cv_k}{S(iv_j, cv_k)} - \min\limits_{iv_j \neq cv_k}{S(iv_j, cv_k)} \\
	\end{split}
	\label{eq:mix-score}
\end{equation}

where $cv_k$ is the existing centroid of $k^{th}$ cluster in centroid set $C^*$.
Compared to the above schema, this schema chooses a good vertex as begin vertex, and guarantee that centroids are separated as much as possible. But other schemes do not take this point into account, and it will cause the selected centroids may belong to the same practical cluster.
Thus we can select $K$ vertices by selecting the vertex with the smallest mix score at the beginning of the clustering algorithm.

\subsection{Vertex Assignment}

After $K$ centroids have been chosen in the $t^{th}$ iteration, we need to assign all the remaining vertices to centroids (or clusters). We test two different schemes for vertex assignment:

\subsubsection{Closest Centroid}
The simple idea of vertex assignment is to assign the vertex $iv_j$ to its closest centroid $cv^* = \argmax_{cv_k^t} S(iv_j, cv_k^t), cv_k^t \in C^*$. If we have good centroids all the time, this schema will work well. However, if the selection of centroids is not good in an iteration, for example, the edge of a cluster or two centroids in the same cluster, the clustering result will develop in a worse direction.

\subsubsection{Dynamic Assignment}
Considering the problem of the closest centroid schema, we develop a dynamic vertex assignment schema. When looking for the most appropriate cluster for a vertex, we consider not only the distance to centroid but also the distance to all assigned vertices in the cluster. When the assignment process has just begun, it gets the same result as the closest centroid schema. As the assignment process processes, it becomes different and better.
Even if a lousy centroid was selected at the last iteration, for example, choosing at the edge of a cluster, we could still assign the right vertices to the centroid, because the center of the gradually generated cluster that we assign vertex to will progressively approach the actual center of the data space.
The vertex order of assignment will affect the effect of the assignment, so the vertex order will be randomly shuffled in each iteration.
Based on above consideration, we will choose the centroid $cv^* = \argmax_{cv_k^t} \frac{1}{\abs{I_k^{t+1}}} \sum_{iv_l \in I_k^{t+1}} S(iv_j, iv_l), cv_k^t \in I_k^t, cv_k^t \in C^*$ for each vertex $iv_j$.

\subsection{Centroid Update}

After assigning all remaining vertices to centroids (or clusters), we need to update new centroids from the assigned cluster. Centroid update is essential for K-Medoids clustering algorithm. A good centroid update schema should make the whole clustering process develop in a good direction. We test two different schemes for centroid update:

\subsubsection{Most Central}
The simple idea of centroid update is to choose the most central vertex as the new centroid.
Before discussing the concept of most central, we will first define the concept of influence score vector $\vec{s}_k$ of vertex $iv_k \in I_j$ which is a vector of length $\abs{I_k}$. The element of influence score vector $\vec{s}_k$ is the influence score between vertex $iv_k \in I_j$ and other vertices $iv_l \in I_j$:

\begin{equation}
	\vec{s}_k = \begin{bmatrix} S(iv_k, iv_1) & S(iv_k, iv_2) & \cdots & S(iv_k, iv_{\abs{I_j}}) \end{bmatrix}
	\label{eq:score-vertor}
\end{equation}

The average influence score vector $\overline{\vec{s}_j}$ of cluster $I_j$ is a vector of length $\abs{I_j}$. The element of average influence score vector $\overline{\vec{s}_j}(k)$ is the average influence score between vertex $iv_k \in I_j$ and other vertices $iv_l \in I_j$:

\begin{equation}
	\overline{\vec{s}_j} = \frac{1}{\abs{I_j}}
	\begin{bmatrix}
		\sum_{iv_l \in I_j} S(iv_1, iv_l) \\
		\sum_{iv_l \in I_j} S(iv_2, iv_l) \\
		\vdots                            \\
		\sum_{iv_l \in I_j} S(iv_{\abs{I_j}}, iv_l))
	\end{bmatrix} ^T
	\label{eq:avg-score-vertor}
\end{equation}
The new centroid $cv_j^{t+1}$ is whose influence score vector is the closest to the average influence score vector. This can be formulated as

\begin{equation}
	cv_j^{t+1} = \argmin_{iv_k \in I_j} \norm{\vec{s}_k - \overline{\vec{s}_j}}
	\label{eq:most-centrally-cnetroid}
\end{equation}

This schema relies heavily on the good assignment of vertices. If the assignment is bad, it will update to a bad centroid. The bad centroid will lead to a bad assignment. The clustering result will become worse and worse.

\subsubsection{Max Objective}

We think a good centroid update schema should make the clustering result develop in the direction of maximizing the objective function.
Thus we will update centroid $cv_j^{t+1}$ which can make objective function maximum from cluster $I_j$. The objective function of a vertex $iv_k \in I_j$ is defined as:

\begin{equation}
	O(iv_k) = \frac{S(iv_k, I_j)}{\frac{1}{K-1} \sum_{l = 1, l \neq j}^K S(iv_k, I_l)}
	\label{eq:objective-function-centroid}
\end{equation}

The numerator of Eq.(\ref{eq:objective-function-centroid}) represents the distance from the cluster to the new centroid, i.e., intra-cluster influence score. The denominator of Eq.(\ref{eq:objective-function-centroid}) captures the distance from other clusters to the new centroid, i.e., inter-cluster influence score. To make the objective function maximum, that is, to make the numerator maximum and make the denominator minimize. The new centroid will achieve a good balance between larger intra-cluster influence score and smaller inter-cluster influence score.

The new centroid $cv_j^{t+1}$ is whose objective function reaches maximum. This can be formulated as

\begin{equation}
	\begin{split}
		cv_j^{t+1} &= \argmax_{iv_k \in I_j} O(iv_k) \\
		& = \argmax_{iv_k \in I_j}
		\frac{S(iv_k, I_j)}{\frac{1}{K-1} \sum_{l = 1, l \neq j}^K S(iv_k, I_l)}
	\end{split}
	\label{eq:max-objective-cnetroid}
\end{equation}

\subsection{Algorithm Summarization}

Given an institution graph $IG$ and $N$ associated influence graphs $FG_i$, Table~\ref{table:algo} shows the steps for partitioning the institution graph into $K$ close collaboration clusters.

\begin{table}[!htbp]
	\caption{{\sc GImpact} Algorithm}
	\label{table:algo}
	\centering
	\hrule
	\vspace{0.25\baselineskip}
	\begin{algorithmic}[1]
		\REQUIRE A institution graph $IG$, $N$ associated influence graphs $FG_i$, the size of cluster $K$, the weights $\alpha, \omega_1, \omega_2, \cdots, \omega_N$
		\ENSURE $K$ clusters $I_1, I_2, \cdots, I_K$
		\STATE Calculate $S_0, S_1, S_2, \cdots, S_N$ respectively
		\STATE Integrate $S_0, S_1, S_2, \cdots, S_N$ into $S$ with weights \\ $\alpha, \omega_1, \omega_2, \cdots, \omega_N$
		\STATE Choose $K$ initial centroids $C^*$
		\REPEAT
		\STATE Assign each vertex $iv \in I$ to the centroid $cv^*$
		\STATE Update new centroid $C_j^{t+1}$
		\UNTIL{Centroids no longer change \OR reach maximum iteration times}
		\RETURN $K$ clusters $I_1, I_2, \cdots, I_K$
	\end{algorithmic}
	\vspace{0.25\baselineskip}
	\hrule
\end{table}

\section{Optimization}

Our research grants dataset contains some special features that we can take advantage of: the discipline attribute contains the hierarchical structure.
The hierarchical structure reveals the relationship between disciplines as supervisory information that can be used by supplementing connections and aggregating connections.
In this section, we will show how we use these special features to optimize the influence analysis.

\subsection{The Hierarchical Structure}

\begin{figure}[!htbp]
	\centering
	\includegraphics[width = 0.4\textwidth]{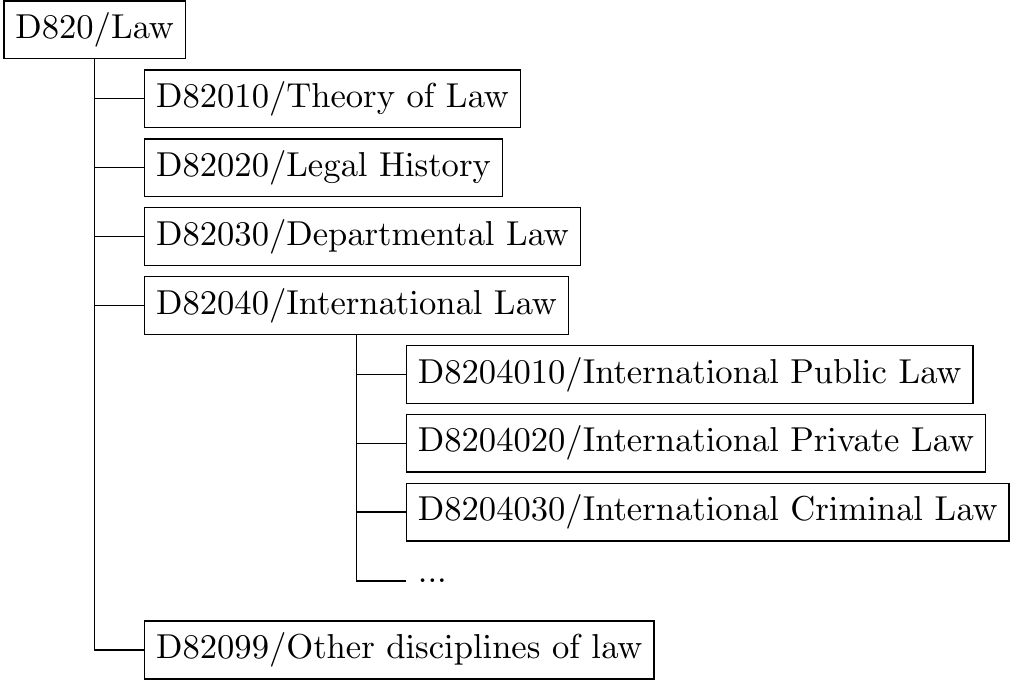}
	\caption{An example fragment of the hierarchical structure of 820/Law}
	\label{fig:tree-disc}
\end{figure}

The discipline attribute has a three-layer structure, which is first-level discipline, second-level discipline, and third-level discipline. A first-level discipline contains multiple second-level disciplines, and a second-level discipline contains multiple third-level disciplines. Figure~\ref{fig:tree-disc} provides an illustrating example fragment of the hierarchical structure of the first-level discipline of D820/Law.

Since the interdisciplinary grants in our dataset are not common, the connections between disciplines are always rare. The hierarchical structure can provide relationships between disciplines to supplement connections. For example, D8204010 (International Public Law) and D8204020 (International Private Law) which are under the same second-level discipline have a close relationship.
Besides, we can treat it as a hand-craft classification. Thus we can classify all 1569 disciplines into 63 first-level discipline categories to aggregate connections if they are under the same first-level discipline. For each keyword, we choose the discipline with the most associated grants as feature discipline which can uniquely represent a keyword. Following discipline classification, we can classify all 20097 keywords into 63 first-level discipline categories as well.

\subsection{Supplement Connections}

Suppose that disciplines under the same discipline have high relevance. We define the coefficient $\lambda_2$ for disciplines under the same second-level discipline, and the coefficient $\lambda_1$ for disciplines under the same first-level discipline. The coefficient $\lambda_2$ should larger than $\lambda_1$, because disciplines under the same second-level discipline should have higher relevance than disciplines under the same first-level discipline.
The supplement weight value is based on the average weight $\overline{w}$ of all edges in the discipline aspect graph. Thus the weight after supplement can be defined as

\begin{equation}
	\tilde{w}_{mn} =  \begin{cases}
		w_{mn} + \lambda_2 * \overline{w} & \textit{same second-level discipline} \\
		w_{mn} + \lambda_1 * \overline{w} & \textit{same first-level discipline}  \\
		w_{mn}                            & \textit{otherwise}
	\end{cases}
	\label{eq:supplement-func}
\end{equation}

\subsection{Aggregate Connections}

The hierarchical structure can be seen as a hand-craft classification. We can apply such classification on influence graph to aggregate connections between institution and aspect attributes. Such aggregation can ignore the differences between particular research directions under the same research area to make co-influence collaboration pattern more visible, and improve computing performance for a large-scale graph.

To apply classification on influence graph, we can partition the influence graph $FG_i$ into $M_i$ parts, denoted by $FP_{1}, FP_{2}, \cdots, FP_{M_i}$. We can construct a $M_i \times (N_{FG_i}+N_{IG})$ auxiliary matrix $\vec{A}_i$ for aggregation. The rows of the auxiliary matrix represent $M_i$ disjoint parts, and the columns of the auxiliary matrix represent institutions and aspect attributes. The auxiliary matrix $\vec{A}_i$ is defined by

\begin{equation}
	\vec{A}_i = \begin{bmatrix}
		p_{11}   & p_{12}   & \cdots & p_{1N_{FG_i}}   & 0      & \cdots & 0      \\
		p_{21}   & p_{22}   & \cdots & p_{2N_{FG_i}}   & 0      & \cdots & 0      \\
		\vdots   & \vdots   & \ddots & \vdots          & \vdots & \ddots & \vdots \\
		p_{M_i1} & p_{M_i2} & \cdots & p_{M_iN_{FG_i}} & 0      & \cdots & 0      \\
	\end{bmatrix}
	\label{eq:aux-matrix}
\end{equation}

where $p_{mn}$ is the probability of aspect attribute $av_n$ belonging to part $FP_{m}$. Then just multiply the auxiliary matrix by multi-hop influence spread kernel, we can get the aggregated influence spread kernel $\vec{A}_i\me^{\alpha t \vec{H}_i}$ which is a $M_i \times (N_{FG_i}+N_{IG})$ matrix. Thus the formulation of the co-influence collaboration pattern $\vec{p}_m$ of institution $iv_m \in I$ will be updated to

\begin{equation}
	\vec{p}_m = \begin{bmatrix}
		\vec{A}_i\me^{\alpha t \vec{H}_i}(1, m + N_{FG_i}) \\
		\vec{A}_i\me^{\alpha t \vec{H}_i}(2, m + N_{FG_i}) \\
		\vdots                                             \\
		\vec{A}_i\me^{\alpha t \vec{H}_i}(M_i, m + N_{FG_i})
	\end{bmatrix}
	\label{eq:behavior-pattern-agg-scalar}
\end{equation}

\begin{figure}[!htbp]
	\subfloat[Co-influence Patterns on Discipline]{
		\includegraphics[width=0.23\textwidth]{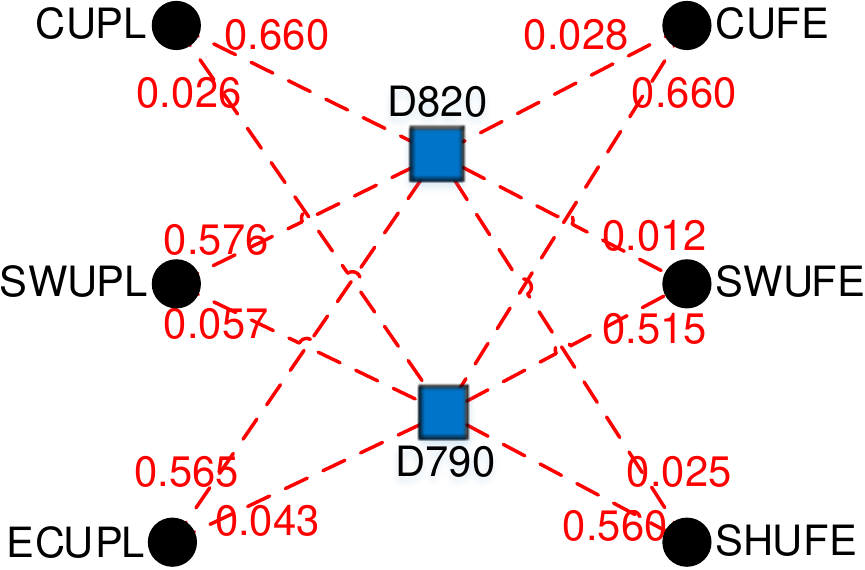}
		\label{fig:pattern-agg-disc}}
	\hfil
	\subfloat[Co-influence Patterns on Keyword]{
		\includegraphics[width=0.23\textwidth]{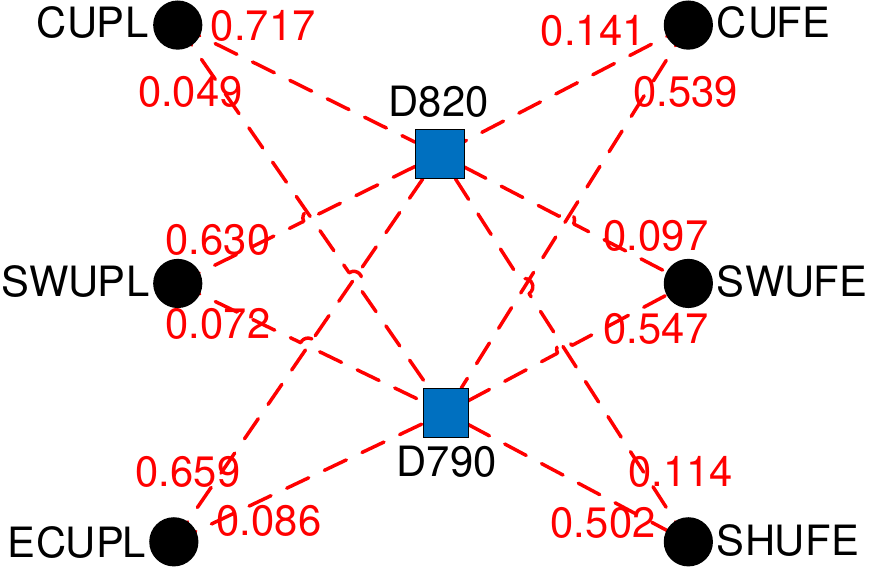}
		\label{fig:pattern-agg-keyword}}
	\caption{Aggregated Result of Co-influence Patterns}
	\label{fig:pattern-agg-dg}
\end{figure}

Figure~\ref{fig:pattern-agg-disc} and Figure~\ref{fig:pattern-agg-keyword} shows examples of aggregated result of multi-hop influence spread kernel of Figure~\ref{fig:framework}e and Figure~\ref{fig:framework}f respectively.
The red numbers on red dash lines measure the scientific influence that discipline categories received from institutions.
As shown in the figures, CUPL, SWUPL, and ECUPL have a very high influence on D820 (Law) research area. Meanwhile, CUFE, SWUFE, and SHUFE have a very high influence on D790 (Economics)research area.
Comparing to the result before aggregation, we ignore the differences under the same first-level discipline to make the similarity of institutions with the same research area much higher. It played a critical role in the graph clustering analysis.

\section{Evaluation}

In this section, we will show the effectiveness of our proposed influence analysis approach {\sc GImpact} and the efficiency of our optimization using a real grants dataset collected from the Social Sciences Management Databases of Chinese Universities (SMDB).

\subsection{Datasets and Experiments Setup}

In our experiments, we choose institution as primary attribution and discipline and keyword as the set of secondary attributes.
We vary the number of clusters K = 25, 50, 100, 200. The associated weights for different K we used in the experiments list at Figure~\ref{tab:weight-full}.
All experiments are conducted on Windows10 with 8GB 1600MHz DDR3 memory and 3.3GHz Intel Core i5. We implement all algorithms in Java 8.

\begin{table}[!htbp]
	\scriptsize
	\renewcommand{\arraystretch}{1.5}
	\caption{The Weights for The Overall Scientific Influence Score}
	\label{tab:weight-full}
	\centering
	\begin{tabular}{
		p{0.225\textwidth}<{\raggedright}
		p{0.035\textwidth}<{\raggedright}
		p{0.035\textwidth}<{\raggedright}
		p{0.035\textwidth}<{\raggedright}
		p{0.035\textwidth}<{\raggedright}}
		\hline
		Type                           & $K$ & $\alpha$ & $w_1$ & $w_2$ \\
		\hline
		Institution                    & 25  & 1        & -     & -     \\
		Institution+Discipline         & 25  & 0.7      & 1.3   & -     \\
		Institution+Keyword            & 25  & 1.3      & 0.7   & -     \\
		Institution+Discipline+Keyword & 25  & 0.15     & 1.9   & 0.95  \\
		Institution                    & 50  & 1        & -     & -     \\
		Institution+Discipline         & 50  & 0.5      & 1.5   & -     \\
		Institution+Keyword            & 50  & 0.25     & 1.75  & -     \\
		Institution+Discipline+Keyword & 50  & 0.8      & 2.1   & 0.1   \\
		Institution                    & 100 & 1        & -     & -     \\
		Institution+Discipline         & 100 & 1.45     & 0.55  & -     \\
		Institution+Keyword            & 100 & 0.25     & 1.75  & -     \\
		Institution+Discipline+Keyword & 100 & 2.05     & 0.8   & 0.15  \\
		Institution                    & 200 & 1        & -     & -     \\
		Institution+Discipline         & 200 & 1.55     & 0.45  & -     \\
		Institution+Keyword            & 200 & 0.65     & 1.35  & -     \\
		Institution+Discipline+Keyword & 200 & 2.2      & 0.6   & 0.2   \\
		\hline
	\end{tabular}
\end{table}

\subsection{Evaluation Methods}

We use two methods to evaluate the quality of the clustering result. The first metric is density of clusters, can be defined as:

\begin{equation}
	density(\{I_j\}^K_{j=1}) = \sum^K_{j=1}
	\frac{|\{(iv_m, iv_n)|iv_m, iv_n \in I_j\}|}{|IE|}
	\label{eq:density}
\end{equation}

Density measures the cohesiveness within clusters. It reflects the consistent between clustering results and current institution collaboration relationships. The larger the density value, the more cohesive the clustering results.

The second metric is the Davies-Bouldin Index (DBI) \cite{davies1979cluster} which measures the uniqueness of clusters.

\begin{equation}
	\sigma_j = \frac{1}{\abs{I_j}} \sum_{iv_l\in I_j} S(cv_j, iv_l)
	\label{eq:avg-score}
\end{equation}

\begin{equation}
	DBI(\{I_j\}^K_{j=1}) = \frac{1}{K} \sum^K_{j=1} max_{k \neq j} \frac{S(cv_j, cv_k)}{\sigma_j + \sigma_k}
	\label{eq:dbi}
\end{equation}

where $cv_x$ is the centroid of $I_x$, $S(cv_j, cv_k)$ is the influence score between vertex $cv_i$ and vertex $cv_j$, $\sigma_x$ is the average influence score of vertices in $I_x$ to $cv_x$. A clustering with higher intra-cluster collaboration score and lower inter-cluster collaboration score will have a lower DBI value.

\subsection{Model Effectiveness}

\begin{figure}[!htbp]
	\centering
	\subfloat[Density]{
		\includegraphics[width=0.23\textwidth]{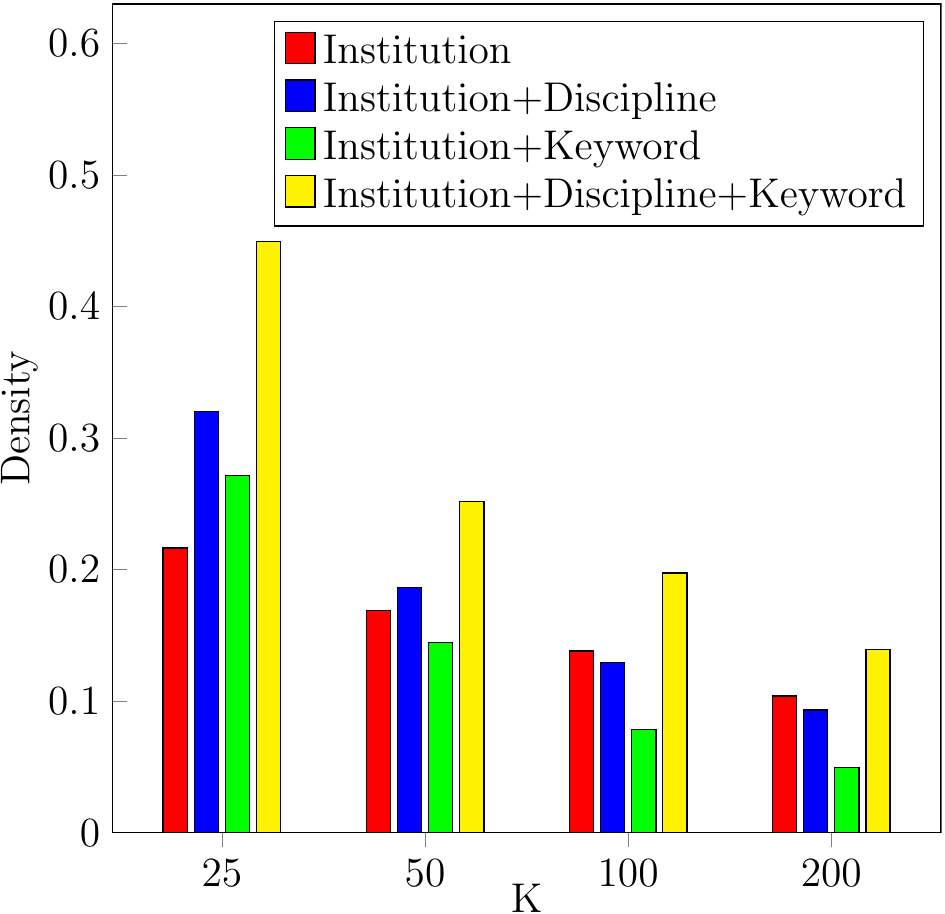}
		\label{fig:quality-full-density}
	}
	\hfil
	\subfloat[DBI]{
		\includegraphics[width=0.23\textwidth]{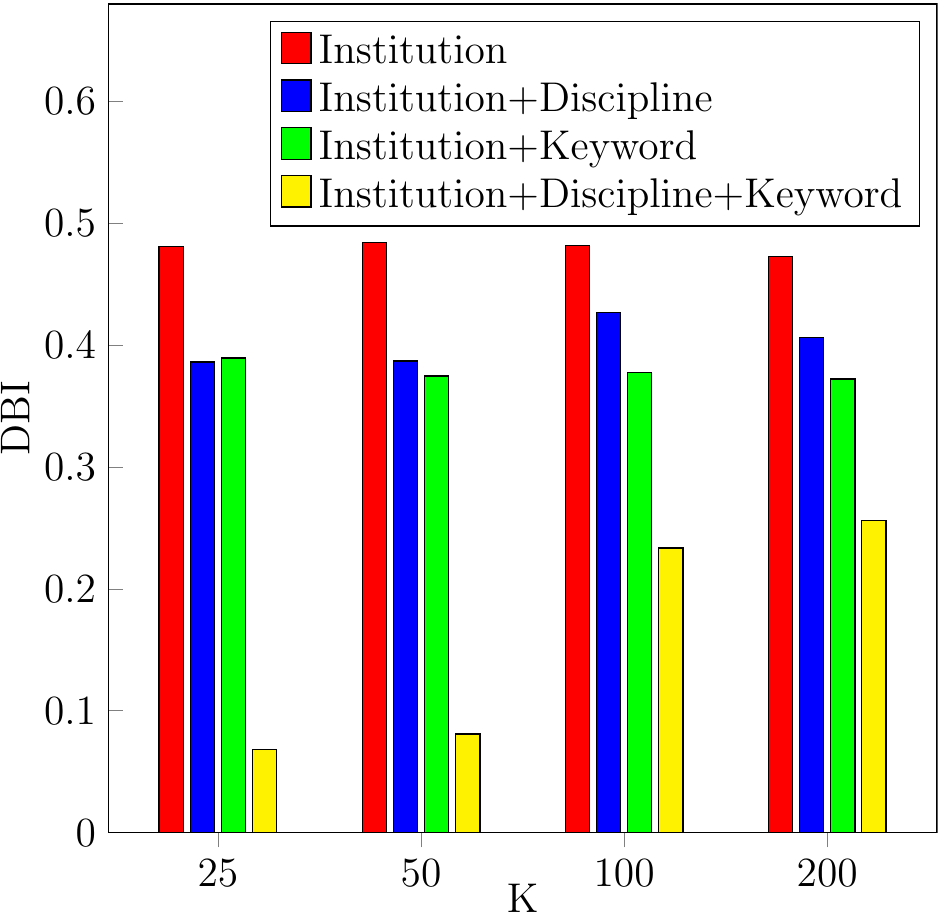}
		\label{fig:quality-full-dbi}
	}
	\caption{Different Influence Graph Comparison}
	\label{fig:quality-full}
\end{figure}

We will evaluate the effectiveness of the overall scientific influence score by using a different number of influence graphs under different K. By comparing the clustering results using different influence graphs, we can learn whether integrating multiple influence graphs has effectiveness on clustering analysis.
Figure~\ref{fig:quality-full-density} and Figure~\ref{fig:quality-full-dbi} show the density comparison and the DBI comparison respectively by varying the number of clusters K = 25, 50, 100, 200.

The clustering results with both discipline influence graph and keyword influence graph have a significantly higher value of density and a substantially lower value of DBI. It indicates that the overall scientific influence score improves the quality of clustering effectively.
It is because the overall scientific influence score considers not only self-influence collaboration patterns but also co-influence collaboration patterns. The clustering results can capture the full feature of the whole research grants network rather than the simple feature only from the institution graph.

Meanwhile, the clustering results with only either discipline influence graph or keyword influence graph are not as good as the clustering results with both two influence graphs. It indicates that just either discipline or keyword cannot describe the research area comprehensively. When K is increasing, the clustering results are even worse than not using influence graphs.

\subsection{Clustering Comparison}

We will evaluate the comparison between different schemes in the clustering analysis under different K. By comparing the clustering results using different schemes, we can evaluate whether good or bad the different schemes are.

\subsubsection{Centroid Initialization Comparison}

\begin{figure}[!htbp]
	\centering
	\subfloat[Density]{
		\includegraphics[width=0.23\textwidth]{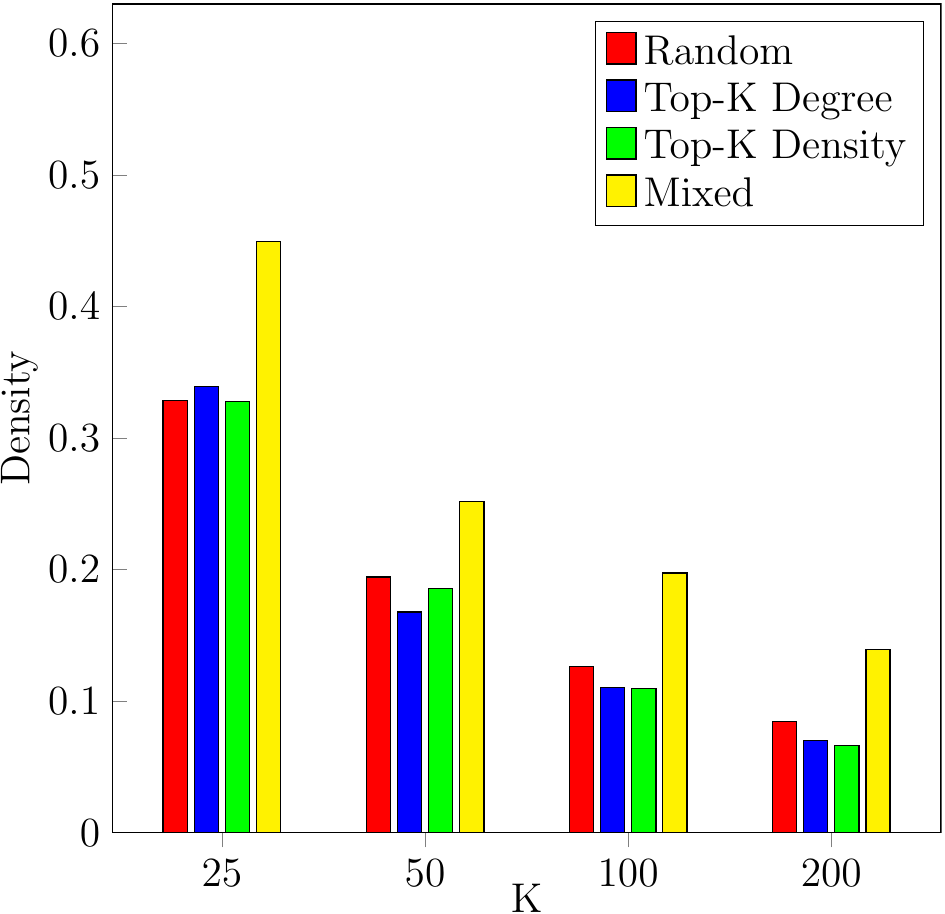}
		\label{fig:cluster-init-full-density}
	}
	\hfil
	\subfloat[DBI]{
		\includegraphics[width=0.23\textwidth]{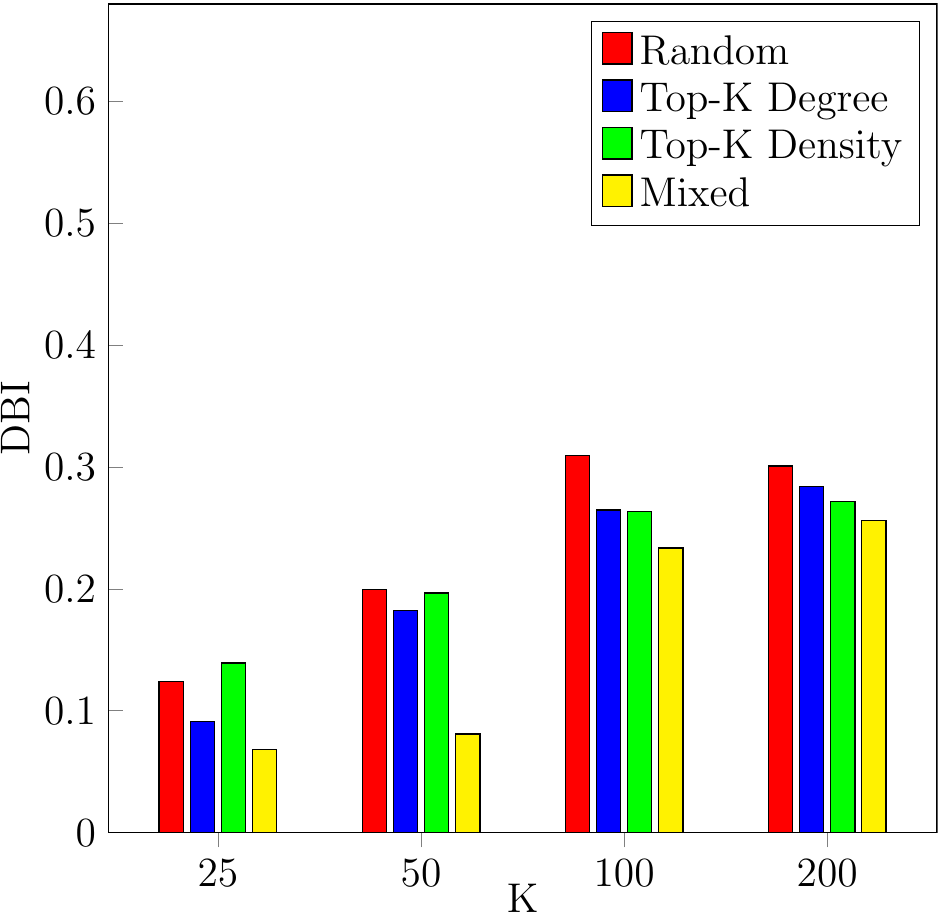}
		\label{fig:cluster-init-full-dbi}
	}
	\caption{Centroid Initialization Comparison}
	\label{fig:cluster-init-full}
\end{figure}

Figure~\ref{fig:cluster-init-full-density} and Figure~\ref{fig:cluster-init-full-dbi} show the quality comparison with different initialization schemes discussed above.
The clustering results with mixed schema have significantly better quality than other schemes. It is because the mixed schema guarantee that initialized centroids are separated as much as possible, while other schemes do not take it into account. If bad centroids are chosen, it is hard to bring it back to the good result. Besides, the top-K degree schema and the top-K density schema are slightly better than the random schema.

\subsubsection{Vertex Assignment Comparison}

\begin{figure}[!htbp]
	\centering
	\subfloat[Density]{
		\includegraphics[width=0.23\textwidth]{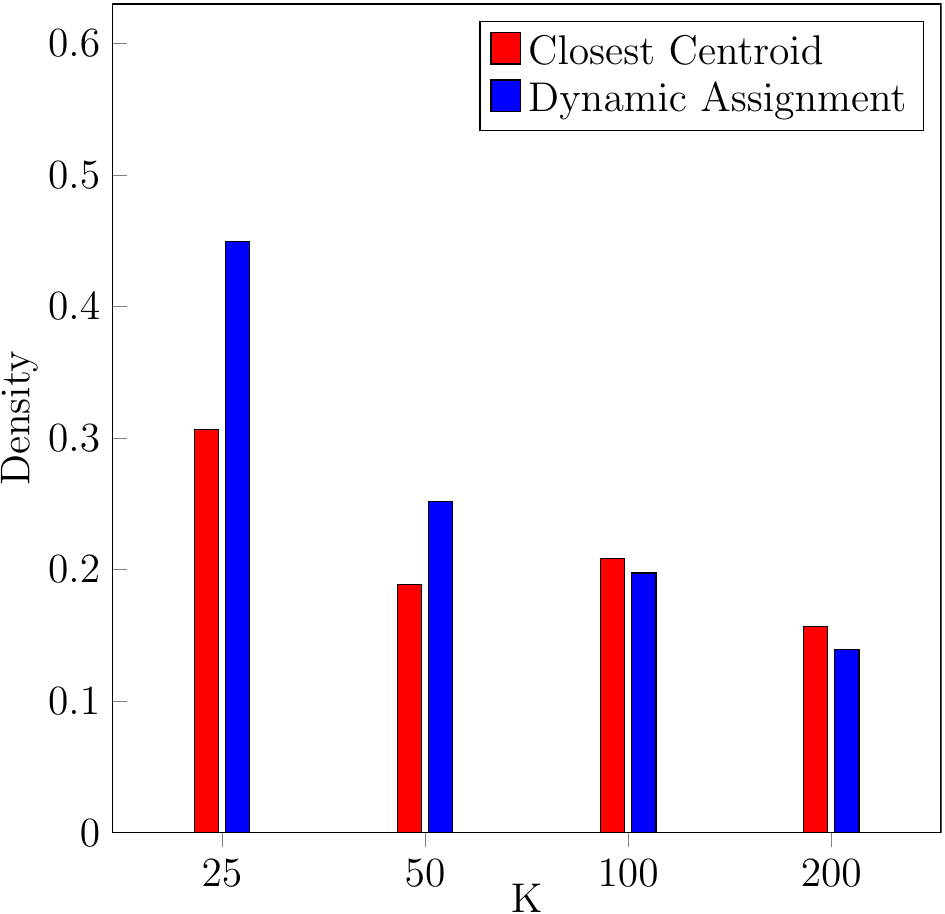}
		\label{fig:cluster-assign-full-density}
	}
	\hfil
	\subfloat[DBI]{
		\includegraphics[width=0.23\textwidth]{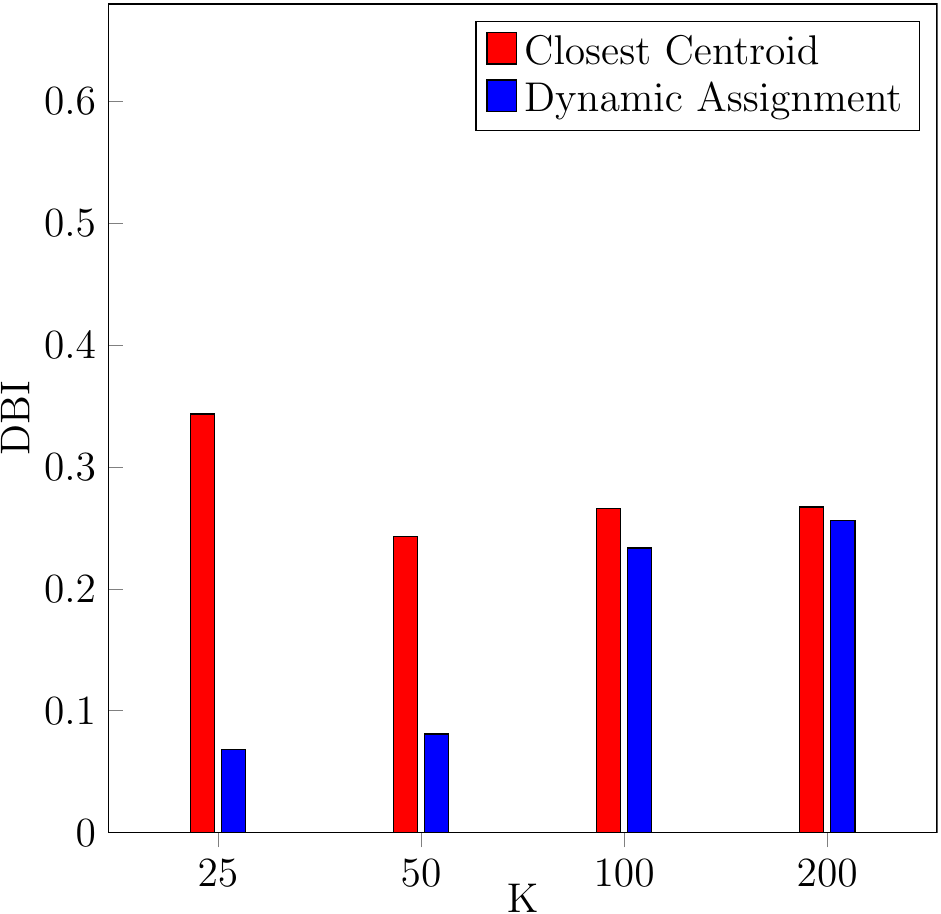}
		\label{fig:cluster-assign-full-dbi}
	}
	\caption{Vertex Assignment Comparison}
	\label{fig:cluster-assign-full}
\end{figure}

Figure~\ref{fig:cluster-assign-full-density} and Figure~\ref{fig:cluster-assign-full-dbi} show the quality comparison with different vertex assignment schemes discussed above.
When K is small, the clustering results with dynamic assignment schema have significantly better quality than the closest centroid schema. Moreover, when K is large, the clustering results with dynamic assignment schema and closest centroid schema have almost the same quality.
It is because when K is small, the number of elements in the cluster is more than the situation that K is large. The more elements in the cluster, the effectiveness of dynamic assignment schema will be more visible. When the number of elements is relatively small, i.e., K is relatively large, the quality of these two schemes should be nearly the same.

\subsubsection{Centroid Update Comparison}

\begin{figure}[!htbp]
	\centering
	\subfloat[Density]{
		\includegraphics[width=0.23\textwidth]{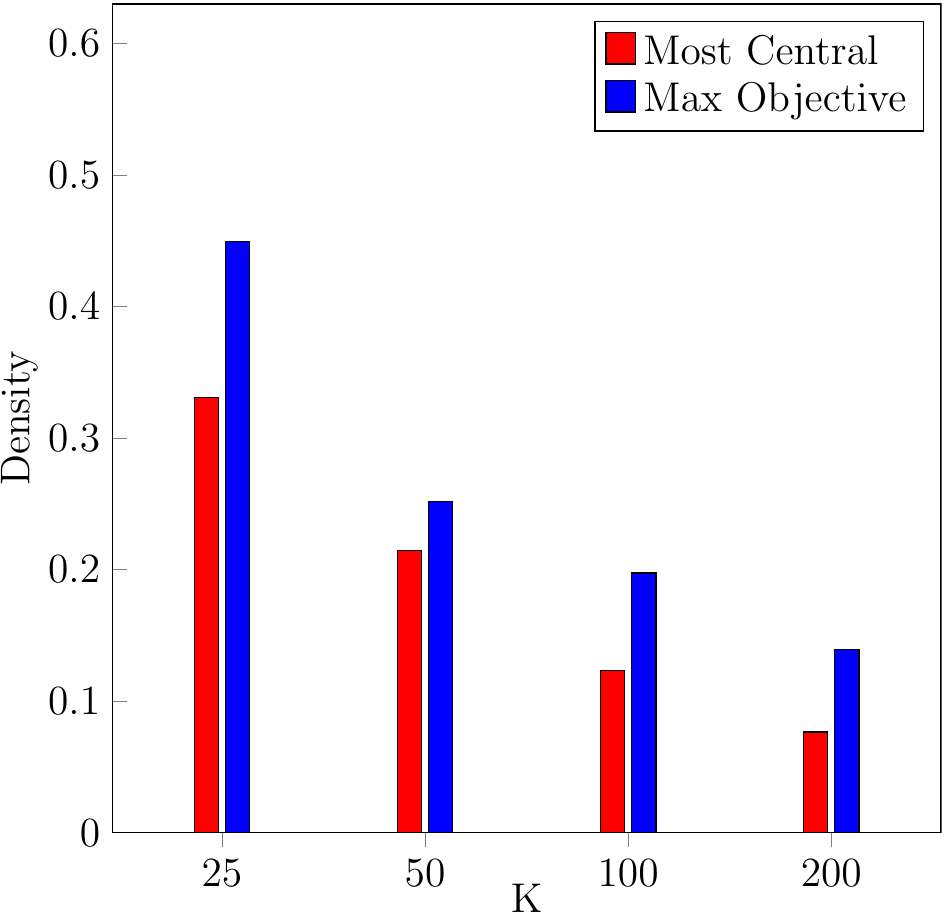}
		\label{fig:cluster-update-full-density}
	}
	\hfil
	\subfloat[DBI]{
		\includegraphics[width=0.23\textwidth]{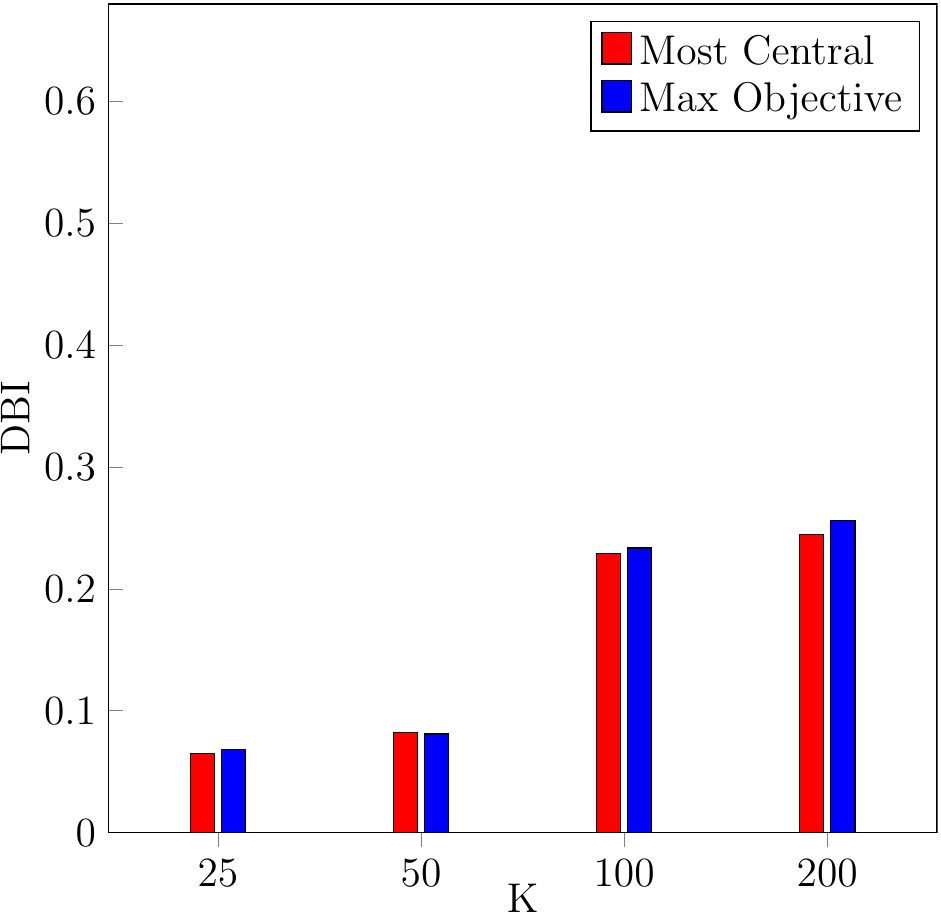}
		\label{fig:cluster-update-full-dbi}
	}
	\caption{Centroid Update Comparison}
	\label{fig:cluster-update-full}
\end{figure}

Figure~\ref{fig:cluster-update-full-density} and Figure~\ref{fig:cluster-update-full-dbi} show the quality comparison with different centroid update schemes discussed above.
The clustering results with max objective schema have a significantly higher density value than the most central schema, and the DBI of these two schemes is nearly the same. It demonstrates that max objective schema can find better centroid than the most central schema.
It is because the max objective schema can achieve a good balance between larger intra-cluster influence score and smaller inter-cluster influence score.

\subsection{Optimization Efficiency}

We will evaluate the quality comparison between different optimization strategies under different K. By comparing different optimization strategies, we can evaluate the pros and cons of different optimization strategies.

\begin{figure}[!htbp]
	\centering
	\subfloat[Density]{
		\includegraphics[width=0.23\textwidth]{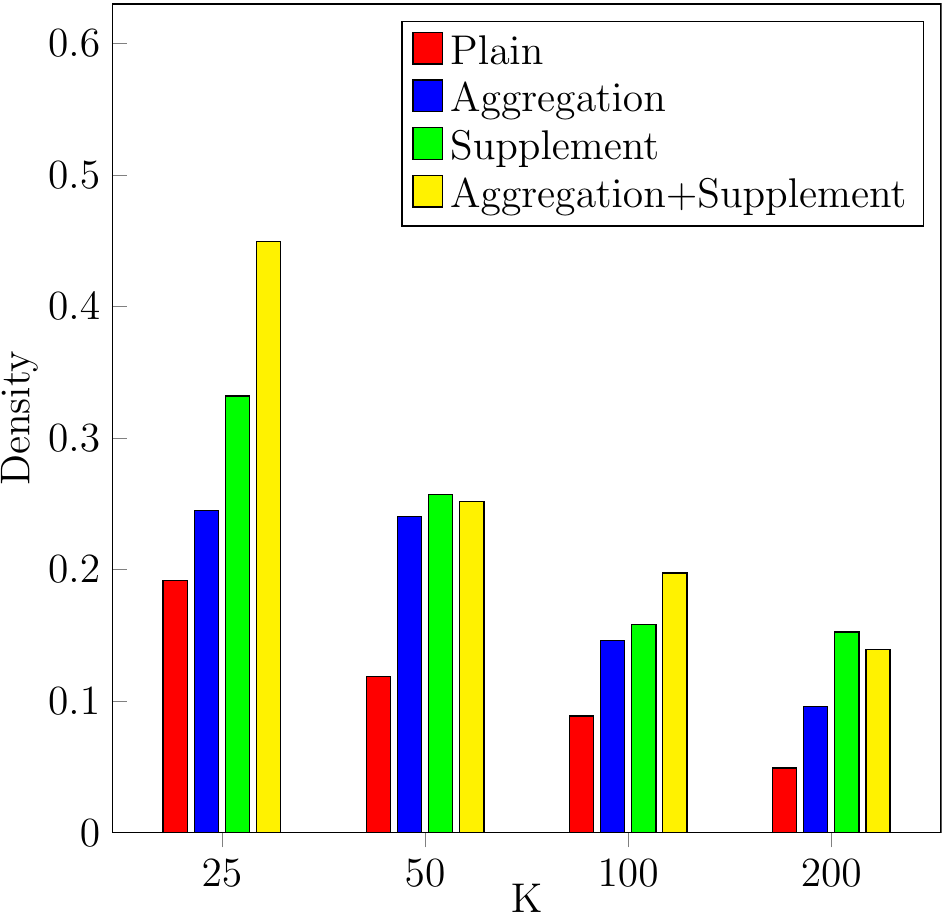}
		\label{fig:opti-hier-full-density}
	}
	\hfil
	\subfloat[DBI]{
		\includegraphics[width=0.23\textwidth]{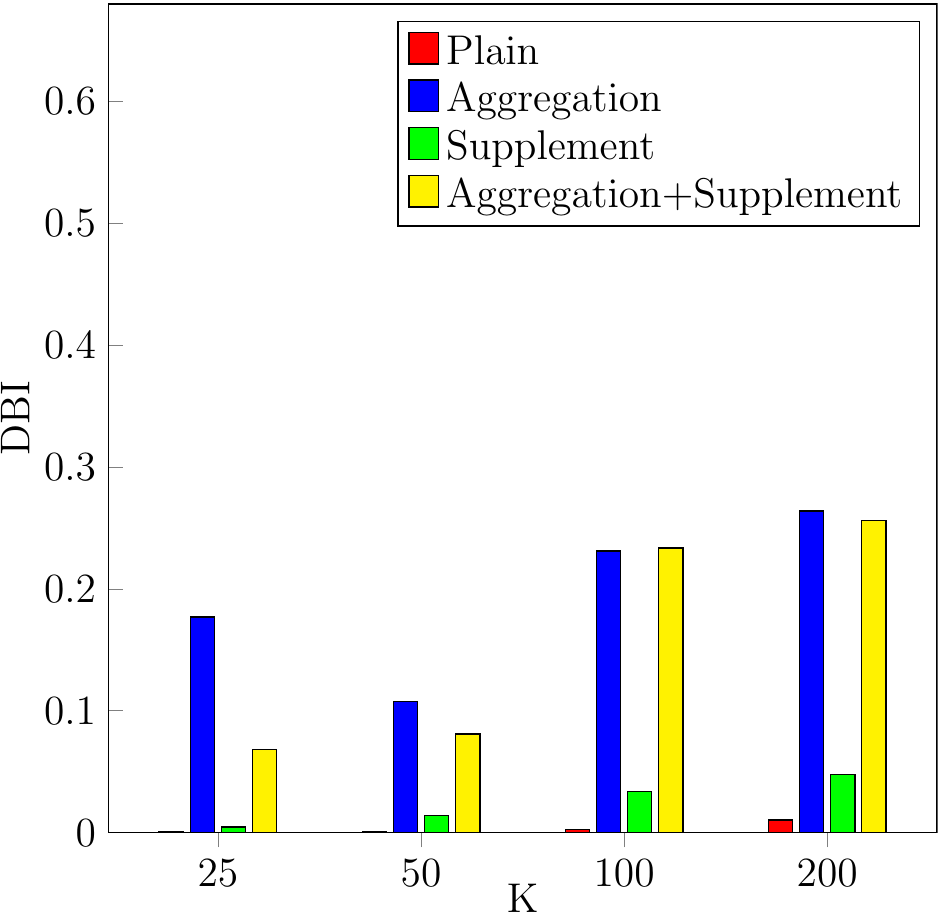}
		\label{fig:opti-hier-full-dbi}
	}
	\caption{Hierarchy Optimization Comparison}
	\label{fig:opti-hier-full}
\end{figure}

Figure~\ref{fig:opti-hier-full-density} and Figure~\ref{fig:opti-hier-full-dbi} show the quality comparison of different optimization strategies discussed above.
As shown in the figure, both supplement and aggregation can increase the density value, the combination of aggregation and supplement will have better results. However, the aggregation will increase the DBI value notably, because the order of magnitude of the influence score will be larger after aggregation. It will make the DBI value increasing, but the density value will not be affected by the order of magnitude of influence score.

\subsection{Case Study}

\subsubsection{Co-influence Collaboration Patterns}

Table~\ref{tab:disc-pattern} and Table~\ref{tab:keyword-pattern} show details of co-influence collaboration patterns of discipline and keyword after aggregation. For each row, we can learn the probability that an institution belongs to a first-level discipline category. For each column, we can learn the contribution of an institution to a certain first-level discipline category, i.e., the institution-level leadership in different research subject areas.

From these two tables, we can learn some interesting phenomenon.
First, comprehensive institutions, e.g., \underline{PKU}, \underline{RUC}, and \underline{WHU}, usually have relatively high values in all research subject areas.
Meanwhile featured institutions, e.g., \textbf{CUFE}, \textbf{CUPL}, and \textbf{BNU}, usually have very high scores in their primary research subject areas and have relatively low scores in other research subject areas. It is very consistent with our common sense.
Second, by summing each column, we can learn which disciplines are the primary research subject areas of social science.
Meanwhile, by summing each row, we can learn which institutions have more contribution to social science development.

\begin{table}[!htbp]
	\scriptsize
	\renewcommand{\arraystretch}{1.5}
	\caption{The Co-influence Patterns of Discipline}
	\label{tab:disc-pattern}
	\centering
	\begin{tabular}{
		p{0.072\textwidth}<{\raggedright}
		p{0.04\textwidth}<{\raggedright}
		p{0.04\textwidth}<{\raggedright}
		p{0.04\textwidth}<{\raggedright}
		p{0.04\textwidth}<{\raggedright}
		p{0.04\textwidth}<{\raggedright}
		p{0.04\textwidth}<{\raggedright}}
		\hline
		Institution     & D190           & D630  & D740  & D790           & D820           & D870  \\
		\hline
		\underline{PKU} & 0.015          & 0.019 & 0.086 & 0.123          & 0.061          & 0.059 \\
		\underline{RUC} & 0.017          & 0.077 & 0.023 & 0.254          & 0.108          & 0.081 \\
		\textbf{BNU}    & \textbf{0.297} & 0.015 & 0.093 & 0.048          & 0.035          & 0.022 \\
		\textbf{CUFE}   & 0.022          & 0.055 & 0.008 & \textbf{0.479} & 0.028          & 0.011 \\
		\textbf{CUPL}   & 0.127          & 0.009 & 0.031 & 0.026          & \textbf{0.660} & 0.006 \\
		FUDAN           & 0.025          & 0.023 & 0.093 & 0.117          & 0.016          & 0.056 \\
		ECNU            & 0.092          & 0.022 & 0.070 & 0.076          & 0.014          & 0.011 \\
		SHUFE           & 0.007          & 0.065 & 0.016 & 0.560          & 0.025          & 0.006 \\
		ECUPL           & 0.011          & 0.021 & 0.021 & 0.043          & 0.565          & 0.011 \\
		\underline{WHU} & 0.009          & 0.052 & 0.020 & 0.115          & 0.106          & 0.214 \\
		SWUFE           & 0.010          & 0.070 & 0.010 & 0.515          & 0.012          & 0.010 \\
		SWUPL           & 0.012          & 0.028 & 0.022 & 0.057          & 0.576          & 0.009 \\
		\hline
	\end{tabular}
\end{table}

\begin{table}[!htbp]
	\scriptsize
	\renewcommand{\arraystretch}{1.5}
	\caption{The Co-influence Patterns of Keyword}
	\label{tab:keyword-pattern}
	\centering
	\begin{tabular}{
		p{0.072\textwidth}<{\raggedright}
		p{0.04\textwidth}<{\raggedright}
		p{0.04\textwidth}<{\raggedright}
		p{0.04\textwidth}<{\raggedright}
		p{0.04\textwidth}<{\raggedright}
		p{0.04\textwidth}<{\raggedright}
		p{0.04\textwidth}<{\raggedright}}
		\hline
		Institution     & D190           & D630  & D740  & D790           & D820           & D870  \\
		\hline
		\underline{PKU} & 0.004          & 0.065 & 0.207 & 0.122          & 0.206          & 0.015 \\
		\underline{RUC} & 0.029          & 0.149 & 0.092 & 0.161          & 0.197          & 0.015 \\
		\textbf{BNU}    & \textbf{0.060} & 0.022 & 0.144 & 0.041          & 0.178          & 0.016 \\
		\textbf{CUFE}   & 0.008          & 0.152 & 0.022 & \textbf{0.539} & 0.140          & 0.008 \\
		\textbf{CUPL}   & 0.008          & 0.027 & 0.045 & 0.049          & \textbf{0.717} & 0.003 \\
		FUDAN           & 0.007          & 0.077 & 0.226 & 0.172          & 0.075          & 0.013 \\
		ECNU            & 0.050          & 0.040 & 0.117 & 0.096          & 0.055          & 0.009 \\
		SHUFE           & 0.003          & 0.118 & 0.096 & 0.502          & 0.114          & 0.001 \\
		ECUPL           & 0.004          & 0.024 & 0.048 & 0.086          & 0.659          & 0.003 \\
		\underline{WHU} & 0.004          & 0.110 & 0.100 & 0.183          & 0.190          & 0.048 \\
		SWUFE           & 0.003          & 0.171 & 0.086 & 0.547          & 0.097          & 0.002 \\
		SWUPL           & 0.004          & 0.049 & 0.050 & 0.072          & 0.630          & 0.003 \\
		\hline
	\end{tabular}
\end{table}

\subsubsection{Influence Scores}

Table~\ref{tab:self-influence} shows self-influence scores of PKU, FUDAN, CUFE, SHUFE, CUPL, and ECUPL. The influence matrix is a symmetric matrix. Except for the diagonal, we highlight the self-influence scores higher than 0.003.
We can notice that the self-influence score between institutions is mainly based on geographical position and research area. For example, PKU, CUFE, and CUPL are all located in Beijing, and FUDAN, SHUFE, and ECUPL are all located in Shanghai too.
Meanwhile, the influence score between CUPL and ECUPL achieve 0.005, because they are all study political science and law primarily.

\begin{table}[!htbp]
	\scriptsize
	\renewcommand{\arraystretch}{1.5}
	\caption{Self-influence Score}
	\label{tab:self-influence}
	\centering
	\begin{tabular}{
		p{0.062\textwidth}<{\raggedright}
		p{0.037\textwidth}<{\raggedright}
		p{0.05\textwidth}<{\raggedright}
		p{0.037\textwidth}<{\raggedright}
		p{0.044\textwidth}<{\raggedright}
		p{0.037\textwidth}<{\raggedright}
		p{0.044\textwidth}<{\raggedright}}
		\hline
		Institution & PKU            & FUDAN          & CUFE           & SHUFE          & CUPL           & ECUPL          \\
		\hline
		PKU         & 1.000          & 0.002          & \textbf{0.006} & 0.001          & \textbf{0.004} & 0.002          \\
		FUDAN       & 0.002          & 1.000          & 0.001          & \textbf{0.007} & 0.001          & \textbf{0.004} \\
		CUFE        & \textbf{0.006} & 0.001          & 1.000          & \textbf{0.003} & \textbf{0.008} & \textbf{0.004} \\
		SHUFE       & 0.001          & \textbf{0.007} & \textbf{0.003} & 1.000          & 0.002          & \textbf{0.006} \\
		CUPL        & \textbf{0.004} & 0.001          & \textbf{0.008} & 0.002          & 1.000          & \textbf{0.005} \\
		ECUPL       & 0.002          & \textbf{0.004} & \textbf{0.004} & \textbf{0.006} & \textbf{0.005} & 1.000          \\
		\hline
	\end{tabular}
\end{table}

Table~\ref{tab:co-influence-disc} and Table~\ref{tab:co-influence-keyword} show co-influence scores of discipline and keyword respectively. We highlight the co-influence scores higher than 0.5 expect for the diagonal.
The research area of an institution mainly decided the co-influence score because discipline and keyword can adequately describe the research area of an institution. For instance, the co-influence score of discipline between CUFE and SHUFE can reach 0.935, and the co-influence score of the keyword is 0.807 because they are all study finance and economics primarily. Also, the political institutions of CUPL and ECUPL have a similar result.
Besides, the comprehensive institutions of PKU and FUDAN have high co-influence score as well. It is indicated that the co-influence score can capture the commonality of the research area of institutions.

\begin{table}[!htbp]
	\scriptsize
	\renewcommand{\arraystretch}{1.5}
	\caption{Co-influence Score of Discipline}
	\label{tab:co-influence-disc}
	\centering
	\begin{tabular}{
		p{0.062\textwidth}<{\raggedright}
		p{0.04\textwidth}<{\raggedright}
		p{0.05\textwidth}<{\raggedright}
		p{0.04\textwidth}<{\raggedright}
		p{0.044\textwidth}<{\raggedright}
		p{0.04\textwidth}<{\raggedright}
		p{0.044\textwidth}<{\raggedright}}
		\hline
		Institution & PKU            & FUDAN          & CUFE           & SHUFE          & CUPL           & ECUPL          \\
		\hline
		PKU         & 1.000          & \textbf{0.696} & 0.400          & 0.389          & 0.134          & 0.160          \\
		FUDAN       & \textbf{0.696} & 1.000          & 0.344          & 0.347          & 0.063          & 0.074          \\
		CUFE        & 0.400          & 0.344          & 1.000          & \textbf{0.935} & 0.061          & 0.068          \\
		SHUFE       & 0.389          & 0.347          & \textbf{0.935} & 1.000          & 0.050          & 0.055          \\
		CUPL        & 0.134          & 0.063          & 0.061          & 0.050          & 1.000          & \textbf{0.854} \\
		ECUPL       & 0.160          & 0.074          & 0.068          & 0.055          & \textbf{0.854} & 1.000          \\
		\hline
	\end{tabular}
\end{table}

\begin{table}[!htbp]
	\scriptsize
	\renewcommand{\arraystretch}{1.5}
	\caption{Co-influence Score of Keyword}
	\label{tab:co-influence-keyword}
	\centering
	\begin{tabular}{
		p{0.062\textwidth}<{\raggedright}
		p{0.04\textwidth}<{\raggedright}
		p{0.05\textwidth}<{\raggedright}
		p{0.04\textwidth}<{\raggedright}
		p{0.044\textwidth}<{\raggedright}
		p{0.04\textwidth}<{\raggedright}
		p{0.044\textwidth}<{\raggedright}}
		\hline
		Institution & PKU            & FUDAN          & CUFE           & SHUFE          & CUPL           & ECUPL          \\
		\hline
		PKU         & 1.000          & \textbf{0.732} & 0.297          & 0.415          & 0.281          & 0.307          \\
		FUDAN       & \textbf{0.732} & 1.000          & 0.298          & 0.476          & 0.111          & 0.131          \\
		CUFE        & 0.297          & 0.298          & 1.000          & \textbf{0.807} & 0.157          & 0.198          \\
		SHUFE       & 0.415          & 0.476          & \textbf{0.807} & 1.000          & 0.127          & 0.163          \\
		CUPL        & 0.281          & 0.111          & 0.157          & 0.127          & 1.000          & \textbf{0.948} \\
		ECUPL       & 0.307          & 0.131          & 0.198          & 0.163          & \textbf{0.948} & 1.000          \\
		\hline
	\end{tabular}
\end{table}

Table~\ref{tab:overall-influence} shows the overall scientific influence score when $\alpha$, $\omega_1$, and $\omega_2$ are all set to 1. Except for the diagonal, we highlight the overall scientific influence score higher than 1.
By integrating the self-influence score and the co-influence score, we enhanced the research area part of the self-influence score so that the overall influence score can reflect what we care about by choosing different influence graphs.
Besides, the geographical position part of the self-influence score supplements the research area part of the co-influence score. Thus, we can get an overall description of the scientific influence of institutions.

\begin{table}[!htbp]
	\scriptsize
	\renewcommand{\arraystretch}{1.5}
	\caption{Overall Scientific Influence Score}
	\label{tab:overall-influence}
	\centering
	\begin{tabular}{
		p{0.062\textwidth}<{\raggedright}
		p{0.04\textwidth}<{\raggedright}
		p{0.05\textwidth}<{\raggedright}
		p{0.04\textwidth}<{\raggedright}
		p{0.044\textwidth}<{\raggedright}
		p{0.04\textwidth}<{\raggedright}
		p{0.044\textwidth}<{\raggedright}}
		\hline
		Institution & PKU            & FUDAN          & CUFE           & SHUFE          & CUPL           & ECUPL          \\
		\hline
		PKU         & 3.000          & \textbf{1.429} & 0.703          & 0.806          & 0.419          & 0.470          \\
		FUDAN       & \textbf{1.429} & 3.000          & 0.643          & 0.829          & 0.176          & 0.209          \\
		CUFE        & 0.703          & 0.643          & 3.000          & \textbf{1.745} & 0.226          & 0.269          \\
		SHUFE       & 0.806          & 0.829          & \textbf{1.745} & 3.000          & 0.179          & 0.224          \\
		CUPL        & 0.419          & 0.176          & 0.226          & 0.179          & 3.000          & \textbf{1.808} \\
		ECUPL       & 0.470          & 0.209          & 0.269          & 0.224          & \textbf{1.808} & 3.000          \\
		\hline
	\end{tabular}
\end{table}
\section{Conclusion}

We have presented the design and development of {\sc GImpact}, a grant-based scientific influence analysis service. It takes a graph-theoretic approach to design and develop large scale scientific influence analysis over a large research-grant repository with three original contributions.
First, we mine the grant database to identify and extract important features for grant-based scientific influence analysis and represent such features using graph theoretic models. In our first prototype of {\sc GImpact}, we construct an institution graph and two grant aspect specific influence graphs, i.e., a disciplines graph and a keywords graph.
Second, we utilize the heat-diffusion based influence spread model to calculate the self-influence score and the co-influence scores to compute two types of collaboration relationships.
Third, we compute the overall scientific influence score for every pair of institutions by introducing a weighted sum of the self-influence score and the multiple co-influence scores, and conduct an influence-based clustering analysis. Evaluating {\sc GImpact} using a real grant database, consisting of 2512 institutions and their grants received over a period of 14 years, we show that {\sc GImpact} can effectively identify the grant-based research collaboration groups and provide valuable insight on an in-depth understanding of the scientific influence of research grants on research programs, institution leadership, and future collaboration opportunities.
\ifCLASSOPTIONcompsoc
	\section*{Acknowledgments}
\else
	\section*{Acknowledgment}
\fi

The authors from Huazhong University of Science and Technology, Wuhan, China, are supported by the Chinese university Social sciences Data Center (CSDC) construction projects (2017-2018) from the Ministry of Education, China. The first author, Dr. Yuming Wang, is a visiting scholar at the School of Computer Science, Georgia Institute of Technology, funded by China Scholarship Council (CSC) for the visiting period of one year from December 2017 to December 2018. Prof. Ling Liu's research is partially supported by the USA National Science Foundation CISE grant 1564097 and an IBM faculty award. Any opinions, findings, and conclusions or recommendations expressed in this material are those of the author(s) and do not necessarily reflect the views of the funding agencies.

\ifCLASSOPTIONcaptionsoff
  \newpage
\fi




\bibliographystyle{IEEEtran}
\bibliography{ref/all}




%

\begin{IEEEbiography}[{\includegraphics[width=1in,height=1.25in,clip,keepaspectratio]{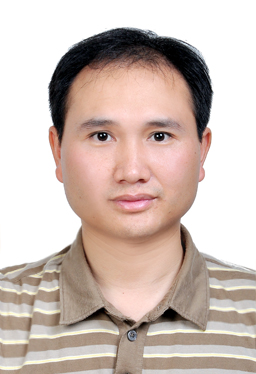}}]{Yuming Wang}
received his B.S. degree in Communication Engineering and Ph.D. degree in Information and Communication Engineering from Huazhong University of Science and Technology (HUST), China, in 2000 and 2005, respectively. He was a Visiting Scholar in the School of Computer Science, College of Computing, Georgia Institute of Technology, for the whole year of 2018. He is currently an Associate Professor in the School of Electronic Information and Communications (EIC), HUST, China. He directs the research programs in Chinese University Social Sciences Data Center (CSDC), HUST, examining various aspects of software service systems and databases for social sciences management and research, including algorithms and applications of data statistics, analysis, mining and visualization. His research interests include Big Data Mining, Knowledge Graph, Machine Learning, Artificial Intelligence and Decision Support.
\end{IEEEbiography}

\begin{IEEEbiography}[{\includegraphics[width=1in,height=1.25in,clip,keepaspectratio]{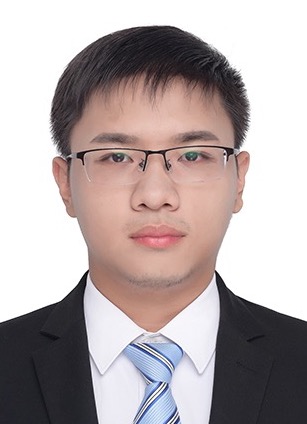}}]{Yanbo Long}
received his B.S. degree in Information and Communication Engineering from the School of Electronic Information and Communications (EIC), Huazhong University of Science and Technology (HUST), China. He is currently working toward the M.S. degree in EIC, HUST. He is also a summer intern of the year 2019 in the branch of Alibaba Cloud in Alibaba Group. His research interests include Big Data Mining, Machine Learning and Scientific Influence Analysis.
\end{IEEEbiography}

\begin{IEEEbiography}[{\includegraphics[width=1in,height=1.25in,clip,keepaspectratio]{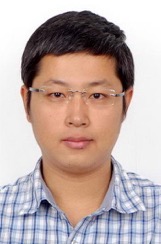}}]{Lai Tu}
received his B.S. degree in Communication Engineering and Ph.D. degree in Information and Communication Engineering from Huazhong University of Science and Technology (HUST), China, in 2002 and 2007, respectively. From 2007 to 2008, he was a Post-Doctoral Fellow in the Department of Eletronics and Information Engineering (EIE), HUST. From 2009 to 2010, he was a Post-Doctoral Researcher in the Department of Computer Science and Information Engineering (CSIE), Nation Cheng Kung University, Taiwan. He is currently an Associate Professor in the School of Electronic and Information and Communications (EIC), HUST. His research interests include Big Data Analysis, Urban Computing, Mobile Computing, and Networking.
\end{IEEEbiography}

\begin{IEEEbiography}[{\includegraphics[width=1in,height=1.25in,clip,keepaspectratio]{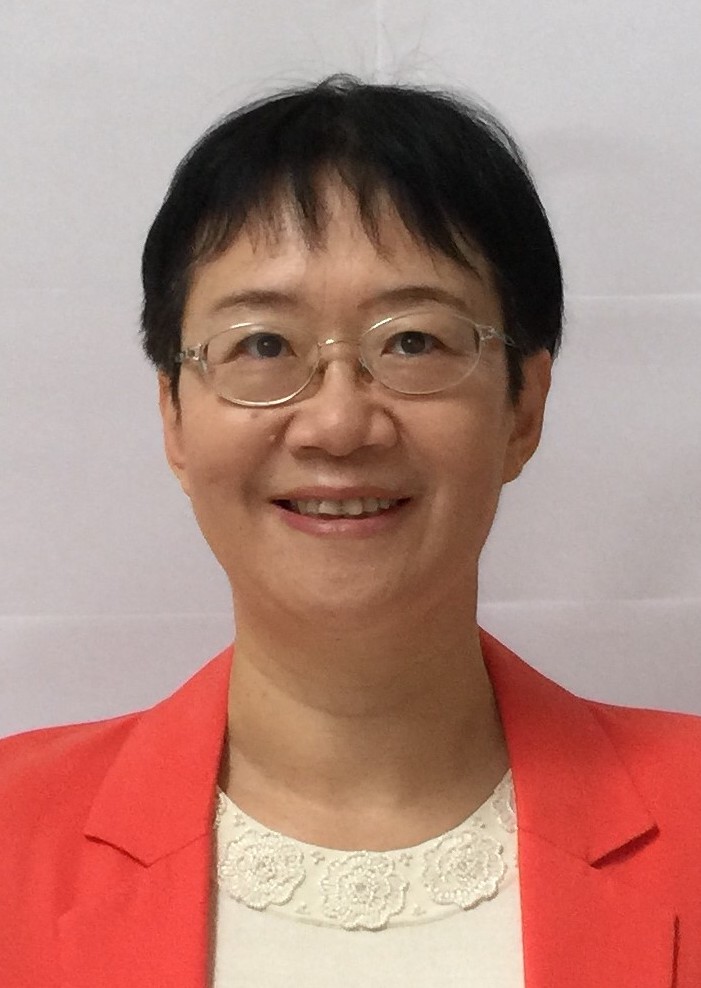}}]{Ling Liu}
is a Professor in the School of Computer Science, College of Computing, Georgia Institute of Technology. She directs the research programs in Distributed Data Intensive Systems Lab (DiSL), examining various aspects of large scale big data systems and analytics, including performance, availability, security, privacy and trust. Prof. Liu is an elected IEEE Fellow and a recipient of the best paper award from numerous top venues, including ICDCS, WWW, IEEE Cloud, IEEE ICWS, ACM/IEEE CCGrid. In addition to serve as general chair and PC chairs of numerous IEEE and ACM conferences in big data, distributed computing, cloud computing, data engineering, very large databases fields, Prof. Liu is currently serving on the editorial board of over a dozen international journals.
\end{IEEEbiography}







\end{document}